# Non-Quasi-Static Effects in Graphene Field-Effect Transistors under High-Frequency Operation

Francisco Pasadas and David Jiménez

*Abstract*— **We investigate the non-quasi-static (NQS) effects in graphene field-effect transistors (GFETs), which are relevant for the device operation at high frequencies as a result of significant carrier inertia. A small-signal NQS model is derived from the analytical solution of drift-diffusion equation coupled with the continuity equation, which can be expressed in terms of modified Bessel functions of the first kind. The NQS model can be conveniently simplified to provide an equivalent circuit of lumped elements ready to be used in standard circuit simulators. Taking into account only first-order NQS effects, accurate GFET based circuit simulations up to several times the cut-off frequency ($f_T$) can be performed. Notably, it reduces to the quasi-static (QS) approach when the operation frequency is below ~ $f_T$/4. To validate the NQS model, we have compared its outcome against simulations based on a multi-segment approach consisting of breaking down the channel length in $N$ equal segments described by the QS model each one.**

## I. INTRODUCTION

GRAPHENE field-effect transistors (GFETs) have been demonstrated to operate within the millimeter-wave range showing intrinsic cut-off frequencies and maximum oscillation frequencies up to hundreds of gigahertz [1],[2]. The design of high-frequency (HF) circuits using these emergent devices requires an appropriate description of its behavior. Many GFET models have been proposed comprising large-signal and small-signal varieties [3-17], but all of them based on a quasi-static (QS) approximation, where the fluctuation of the varying terminal voltages is assumed to be slow, so the stored charge in the device could follow the voltage variations. Such an approximation is found to be valid when the transition time for the voltage to change is larger than the transit time of the carriers from source to drain. This approximation works well in many FET based circuits, although it could fail, especially with long-channel devices operating at high frequencies or when the load capacitance is very small [18],[19]. This can present a serious problem in state-of-the-art GFET-based circuit designs, e.g., in predicting phase margins and stability of wide-band amplifiers [20].

In this paper, we present a small-signal non-quasi-static (NQS) model for the intrinsic part of a four-terminal (4T) GFET. Fur such a purpose, we have used a DC-to-HF methodology [19],[20] to extract a general analytical solution of the drift-diffusion equation coupled with the continuity equation. This solution can be expressed in terms of the modified Bessel functions of the first kind forming the basis of the NQS model. Then, a first-order approximation of the analytical solution based on the intrinsic admittance parameters is provided, which can be transformed into an equivalent circuit with lumped elements, analogous to the one derived for conventional silicon CMOS technology. Such lumped elements can be conveniently described in terms of the graphene chemical potentials at the drain and source edges. Thus, the equivalent circuit can be straightforwardly embedded (for instance, using a description in Verilog-A) in standard circuit simulators, enabling the performance prediction of GFET-based circuits at high frequencies. The assessment as well as the analysis of the region of validity of the complete charge-based QS, first-order NQS and numerical NQS models are carried out for a prototype GFET.

## II. MODEL FORMULATION

Based on the previous charge-based QS model for GFETs presented by Pasadas and Jiménez [7]-[9], the graphene transport charge per unit area under time-varying excitations can be expressed in simple form in terms of the chemical potential, $v_C(x,t)$ (the notation considered can be checked in Appendix A):

$$q_T^{'}(x,t) = \frac{k}{2}\left(v_C^2(x,t) + \alpha\right) \qquad (1)$$

where $q^{'}_T(x,t)$ is given as a function of position $x$ along the channel and time $t$ (positive $x$ direction is from source, $x = 0$, to drain, $x = L$; where $L$ is the channel length as shown in the device's cross-section depicted in Fig. 1A).

Manuscript received February 20, 2020; accepted March 19, 2020. This project has received funding from the European Union's (UE) Horizon 2020 research and innovation programme under grant agreements No GrapheneCore2 785219 and No GrapheneCore3 881603, from Ministerio de Ciencia, Innovación y Universidades (MCIU) and Agencia Estatal de Investigación (AEI) under grant agreement RTI 2018-097876-BC21(MCIU/AEI/FEDER, UE), and project 001-P-0011702-GraphCAT: Comunitat Emergent de grafè a Catalunya, co-funded by Fondo Europeo de Desarrollo Regional (FEDER) within the framework of Programa Operatiu FEDER de Catalunya 2014-2020. (*Corresponding author: Francisco Pasadas.*)

F. Pasadas and D. Jiménez are with the Departament d'Enginyeria Electrònica, Escola d'Enginyeria, Universitat Autònoma de Barcelona, Bellaterra (Barcelona) 08193, Spain (e-mail: francisco.pasadas@uab.es).

Digital Object Identifier: 10.1109/TED.2020.2982840

This article has supplementary material available in the online version of this article provided by the authors. The supplementary material (SM) describes the detailed derivation of the non-quasi-static model of GFETs following a DC-to-HF methodology comprising the zero- and first-order approximations.

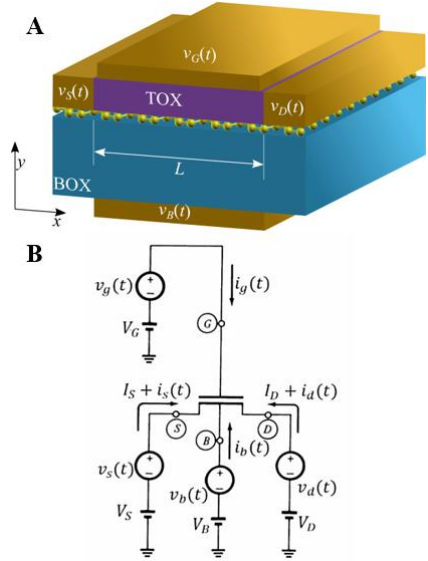

Fig. 1. (A) Cross-section of a four-terminal GFET. A double-gate stack consisting of top and back gate dielectrics (TOX and BOX, respectively) and corresponding metal gates electrostatically control the graphene sheet, which plays the role of the active channel. (B) Schematic of a four-terminal FET operating under small-signal regime showing the terminal DC and AC voltages and currents. To guarantee charge conservation, the sum of the terminal currents must be zero, thus $I_D = -I_S$, where the DC top and back gate currents are considered to be zero $I_B = I_G = 0$; and $i_s(t)+i_b(t)+i_d(t)+i_s(t) = 0$.

In (1), $k = 2q^3/(\pi\hbar^2 v_F^2)$; $q$ is the elementary charge; $\hbar$ is the reduced Planck's constant; $v_F = 3a_{cc}\gamma_0/2\hbar$ is the Fermi velocity; $a_{cc} = 2.49$ Å is the carbon–carbon distance of the honeycomb-like crystal lattice structure [21]; and $\gamma_0 = 3.16$ eV is the interlayer coupling [22]. Moreover, $\alpha = (\pi k_B T/q)^2/3 + 2\sigma_{pud}/k$; $k_B$ is the Boltzmann constant; $T$ is the temperature; and $\sigma_{pud}$ is the residual charge density due to electron–hole puddles [11].

Now, the current in the device is no longer considered the same everywhere along the channel because fast variations are allowed and, thus, should be taken into account. In this regard, we have considered the drift-diffusion mechanism to describe carrier transport coupled with the continuity equation which are given, respectively, by:

$$i_{DS}(x,t) = \mu W q_T(x,t)\left(1 + \frac{k|v_C(x,t)|}{C'}\right)\frac{\partial v_C(x,t)}{\partial x}$$

$$\frac{\partial i_{DS}(x,t)}{\partial x} = W\frac{\partial q_T(x,t)}{\partial t} \rightarrow \frac{\partial i_{DS}(x,t)}{\partial x} = Wk v_C(x,t)\frac{\partial v_C(x,t)}{\partial t}$$

where $i_{DS}(x,t)$ is the time-varying drain-to-source current; $\mu$ is the effective carrier mobility for both electrons and holes (both assumed to be independent of the applied electric field, carrier density or temperature); $W$ is the channel width; and $C'$ = $C'_t + C'_b$, where $C'_t$ ($C'_b$) is the geometrical capacitance per unit area of the top (back) gate oxide shown as TOX (BOX) in Fig. 1A.

Considering that the time-varying excitations are small in amplitude, we can assume that only first-order terms in the AC components are retained. Moreover, if the excitations are sinusoidal, we can describe the AC terms with phasors and, therefore, rewrite (2) in the following form (a detailed explanation of the process followed to get (3) is given in Section A of the Supplementary Material, SM):

$$i = \mu W\frac{k}{2}\left[u\left(\frac{k\alpha}{C'}\frac{|v|}{v} + 2v + 3v\frac{k|v|}{C'}\right)\frac{\partial v}{\partial x} + \left(v^2 + \alpha\right)\left(1 + \frac{k|v|}{C'}\right)\frac{\partial u}{\partial x}\right] \quad (3)$$

$$\frac{\partial i}{\partial x} = j\omega W k v u$$

Applying some transformations to the differential equations describing the small-signal sinusoidal operation of a GFET in (3), the following second-order differential equation has been found to describe the NQS response of such a device:

$$\frac{d^2 i}{dv^2} - \frac{1}{v}\frac{di}{dv} - j\omega\frac{\mu W^2 k^2}{2I_{DS}^2}v^3\left(1 + \frac{k}{C'}|v|\right)i = 0 \quad (4)$$

where $I_{DS}$ is the DC drain current. We have considered that $v^2 \gg \alpha$, implying that the thermal charge density plus electron-hole puddle density is much lower than the charge density produced by the electrostatics (see (1)). That could be the case when the gate bias is far away from the Dirac voltage, which results in unipolar channels. This particular bias point is desirable when the GFET is used as an RF amplifier [23] and will be discussed in depth later.

The second-order differential equation expressed by (4) can be solved in two opposite limits:

- **Case A**: $k|v|/C' \ll 1$

In this case, (4) can be rewritten as follows:

$$\frac{d^2 i}{dv^2} - \frac{1}{v}\frac{di}{dv} - jDv^3 i = 0 \quad (5)$$

where $D = \omega\mu W^2 k^2/(2I_{DS}^2)$. The general solution of (5) is given by:

$$i(v) = v\left[\frac{i_s}{v_{cs}}\frac{n_3(v)p_2 - p_3(v)n_2}{p_2 n_1 - n_2 p_1} + \frac{i_d}{v_{cd}}\frac{p_3(v)n_1 - n_3(v)p_1}{p_2 n_1 - n_2 p_1}\right] \quad (6)$$

with:

$$
\begin{aligned}
n_1 &= I_{-\frac{2}{5}}\left(\frac{2}{5}Dv_{cs}^{5/2}\right) & p_1 &= I_{\frac{2}{5}}\left(\frac{2}{5}Dv_{cs}^{5/2}\right) \\
n_2 &= I_{-\frac{2}{5}}\left(\frac{2}{5}Dv_{cd}^{5/2}\right) & p_2 &= I_{\frac{2}{5}}\left(\frac{2}{5}Dv_{cd}^{5/2}\right) \\
n_3(v) &= I_{-\frac{2}{5}}\left(\frac{2}{5}Dv^{5/2}\right) & p_3(v) &= I_{\frac{2}{5}}\left(\frac{2}{5}Dv^{5/2}\right)
\end{aligned}
\quad (7)
$$

where $D_1 = (-1)^{1/4}D^{1/2}$ and $I_\lambda(z)$ is the modified Bessel function of the first kind [24], where, $\lambda$ is the real order and $z$ is the complex argument. $I_\lambda(z)$ is a convergent series everywhere in the complex $z$-plane. The boundary conditions are given by (see SM Section B to check the analytic computation of $I_\lambda(z)$ and the procedure followed to determine the boundary conditions $i_s$, $i_d$, $v_{cs}$, $v_{cd}$, and $u_{cd}$):

$$i_s = v_{cs}\frac{D_2}{D_1}\frac{u_{cd}\sqrt{v_{cd}}\left[n_1 n_{4s} - p_1 p_{4s}\right] + u_{cs}\sqrt{v_{cs}}\left[p_1 p_{4d} - n_1 n_{4d}\right]}{p_{4d}n_{4s} - n_{4d}p_{4s}}$$

$$i_d = v_{cd}\frac{D_2}{D_1}\frac{u_{cd}\sqrt{v_{cd}}\left[n_2 n_{4s} - p_2 p_{4s}\right] + u_{cs}\sqrt{v_{cs}}\left[p_2 p_{4d} - n_2 n_{4d}\right]}{p_{4d}n_{4s} - n_{4d}p_{4s}}$$

$$v_{cs} = \frac{C' - \sqrt{C'^2 \pm 2k\left[C_t'\left(V_G - V_{G0} - V_S\right) + C_b'\left(V_B - V_{B0} - V_S\right)\right]}}{\pm k}$$

$$v_{cd} = \frac{C' - \sqrt{C'^2 \pm 2k\left[C_t'\left(V_G - V_{G0} - V_D\right) + C_b'\left(V_B - V_{B0} - V_D\right)\right]}}{\pm k}$$

$$u_{cs} = -\frac{C_t'\left(V_g - V_s\right) + C_b'\left(V_b - V_s\right)}{\sqrt{C'^2 \pm 2k\left[C_t'\left(V_G - V_{G0} - V_S\right) + C_b'\left(V_B - V_{B0} - V_S\right)\right]}}$$

$$u_{cd} = -\frac{C_t'\left(V_g - V_d\right) + C_b'\left(V_b - V_d\right)}{\sqrt{C'^2 \pm 2k\left[C_t'\left(V_G - V_{G0} - V_D\right) + C_b'\left(V_B - V_{B0} - V_D\right)\right]}} \quad (8)$$

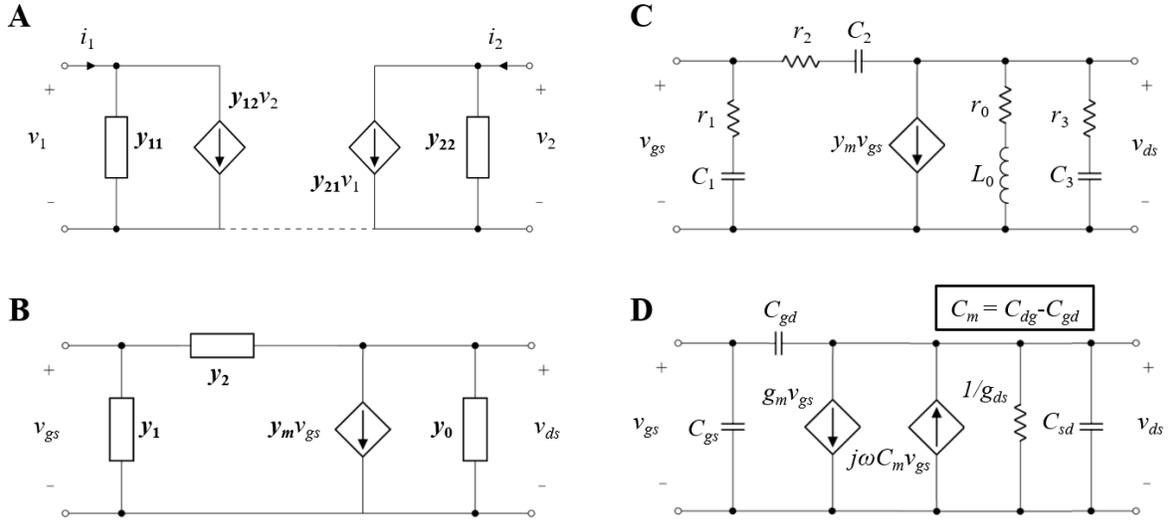

Fig. 2. (A) Equivalent circuit in a general form of a FET in two-port configuration described by the short-circuit admittance parameters. Ports 1 and 2 refer to the gate-source and drain-source ports, respectively. (B) Equivalent circuit of a GFET in two-port configuration in a standard form described by admittance parameters and (C) describing the first-order NQS behavior by means of lumped elements. (D) Complete charge-based QS small-signal model of GFET [27]. The small-signal elements are: $g_m$ transconductance, $g_{ds}$ output conductance and $C_{gs}$, $C_{gb}$, $C_{sd}$ and $C_{dg}$ independent intrinsic capacitances.

with:

$$n_4(v) = I_{-\frac{3}{5}}\left(\frac{2}{5}D_1v^{5/2}\right) \qquad p_4(v) = I_{\frac{3}{5}}\left(\frac{2}{5}D_1v^{5/2}\right) \qquad (9)$$

where $D_2 = jDI_{DS}$; $n_{4s} = n_4(v_{cs})$; $n_{4d} = n_4(v_{cd})$; $p_{4s} = p_4(v_{cs})$; and $p_{4d} = p_4(v_{cd})$. The positive (negative) sign in Equation (8) applies when $C'_t(V_G-V_{G0}-V_X) + C'_b(V_B-V_{B0}-V_X) < 0$ ($>0$), where the subscript $X$ stands for drain ($D$) and source ($S$).

- **Case B:** $k|v|/C' \gg 1$

In this case, (4) can be rewritten as follows:

$$\frac{d^2i}{dv^2} - \frac{1}{v}\frac{di}{dv} - jDv^4i = 0 \qquad (10)$$

where $D = \text{Sign}[v]\omega\mu W^2k^3/(2C'I_{DS}^2)$. The general solution of (10) is also given by (6) but, in this case, we have to consider instead:

$$n_1 = I_{-\frac{1}{3}}\left(\frac{1}{3}D_1v_{cs}^3\right) \qquad p_1 = I_{\frac{1}{3}}\left(\frac{1}{3}D_1v_{cs}^3\right)$$

$$n_2 = I_{-\frac{1}{3}}\left(\frac{1}{3}D_1v_{cd}^3\right) \qquad p_2 = I_{\frac{1}{3}}\left(\frac{1}{3}D_1v_{cd}^3\right) \qquad (11)$$

$$n_3(v) = I_{-\frac{1}{3}}\left(\frac{1}{3}D_1v^3\right) \qquad p_3(v) = I_{\frac{1}{3}}\left(\frac{1}{3}D_1v^3\right)$$

where $D_1 = (-1)^{1/4}D^{1/2}$; $D_{1s} = D_1|_{v=vcs}$; $D_{1d} = D_1|_{v=vcd}$; and the boundary conditions, $i_s$ and $i_d$, are given by:

$$i_s = v_{cs}\frac{D_{1s}D_{2d}u_{cd}v_{cd}[n_1n_{4s}-p_1p_{4s}]+D_{1d}D_{2s}u_{cs}v_{cs}[p_1p_{4d}-n_1n_{4d}]}{D_{1s}D_{1d}[p_{4d}n_{4s}-n_{4d}p_{4s}]}$$

$$i_d = v_{cd}\frac{D_{1s}D_{2d}u_{cd}v_{cd}[n_2n_{4s}-p_2p_{4s}]+D_{1d}D_{2s}u_{cs}v_{cs}[p_2p_{4d}-n_2n_{4d}]}{D_{1s}D_{1d}[p_{4d}n_{4s}-n_{4d}p_{4s}]} \qquad (12)$$

and:

$$n_4(v) = I_{-\frac{2}{3}}\left(\frac{1}{3}D_1v^3\right) \qquad p_4(v) = I_{\frac{2}{3}}\left(\frac{1}{3}D_1v^3\right) \qquad (13)$$

where $D_2 = jDI_{DS}$; $D_{2s} = D_2|_{v=vcs}$; $D_{2d} = D_2|_{v=vcd}$; $n_{4s} = n_4(v_{cs})$; $n_{4d} = n_4(v_{cd})$; $p_{4s} = p_4(v_{cs})$; and $p_{4d} = p_4(v_{cd})$.

In Fig. 1B, the schematic of the terminal currents of a 4T device is shown. In (8) and (12), expressions for computing the sinusoidal AC drain and source terminal currents are provided. To guarantee charge-conservation the sum of the AC terminal currents entering into a device must be zero, thus, $i_g(t)+i_b(t) = -i_d(t)-i_s(t)$. As the top and back gate AC terminal currents are related to the top and back gate geometrical capacitances per unit area [9], the NQS AC terminal currents can be expressed as follows (see SM Subsection B4 for further details):

$$I_s(\omega) = -I_{ds}(0,\omega) = -i_s$$

$$I_d(\omega) = I_{ds}(L,\omega) = i_d$$

$$i_g = I_g(\omega) = -\frac{C'_t}{C'}(i_d - i_s) \qquad (14)$$

$$i_b = I_b(\omega) = -\frac{C'_b}{C'}(i_d - i_s)$$

### A. NQS short-circuit admittance parameters of GFETs

An equivalent circuit based on the short-circuit admittance parameters can be provided by relating the terminal current phasors in (14) to the terminal voltage phasors. They can be easily calculated using the definition, taking the source as reference [20]:

$$y_{jk} = \frac{i_j}{u_{ks}}\bigg|_{u_{ls}=0,l\neq k} \qquad (15)$$

where the subscripts $j$, $k$, and $l$ stand for drain ($d$), top gate ($g$), and back gate ($b$). Only nine admittances out of sixteen are independent, and hence enough to completely characterize the device [19]. The $y$-parameters are important since their design of high-frequency amplifiers is easily done in terms of these parameters. However, in such designs is usual to consider that the back-gate terminal is also AC short-circuited to the source, thus, forming a two-port network in common-source configuration and resulting in the general form of the equivalent circuit depicted in Fig. 2A, which is going to be considered in the following derivation.

We have focused on the Case B (see SM Section C to find the full procedure followed to get the expressions of the admittance parameters for both cases A and B). The procedure begins by rewriting (6) as follows [25]:

$$i(v) = k_1 g_1(v) + k_2 g_2(v) \qquad (16)$$

where:

$$
\begin{aligned}
k_1 &= \frac{i_d n_1 v_{cs} - i_s n_2 v_{cd}}{v_{cs} v_{cd} \left( p_2 n_1 - n_2 p_1 \right)} \left( \frac{D_1}{6} \right)^{\frac{1}{3}} \frac{1}{\Gamma(4/3)} \\
k_2 &= \frac{i_s p_2 v_{cd} - i_d p_1 v_{cs}}{v_{cs} v_{cd} \left( p_2 n_1 - n_2 p_1 \right)} \left( \frac{D_1}{6} \right)^{-\frac{1}{3}} \frac{1}{\Gamma(2/3)} \\
g_1(v) &= v^2 \left( 1 + j \frac{Dv^6}{48} - j \frac{D^3 v^{12}}{8064} + \ldots \right) \\
g_2(v) &= v^2 \left( 1 + j \frac{Dv^6}{24} - j \frac{D^2 v^{12}}{1920} + \ldots \right)
\end{aligned}
\qquad (17)
$$

where $\Gamma$ stands for the Gamma function [24]. If (16) is differentiated, the following equation results:

$$\frac{u}{B} = k_1 F_1(v) + k_2 F_2(v) \qquad (18)$$

where $B = 1/D_2$ and:

$$
\begin{aligned}
F_1(v) &= \frac{dg_1}{dv} \Big/ v^4 \\
F_2(v) &= \frac{dg_2}{dv} \Big/ v^4
\end{aligned}
\qquad (19)
$$

Following the procedure proposed elsewhere [25] for building the NQS equivalent circuit for conventional silicon based MOSFETs, from (16)-(19), the $y$-parameter matrix for the equivalent circuit in Fig. 2A can be written as:

$$
\begin{pmatrix} i_1 \\ i_2 \end{pmatrix} = \begin{pmatrix} y_{11} & y_{12} \\ y_{21} & y_{22} \end{pmatrix} \begin{pmatrix} v_1 \\ v_2 \end{pmatrix}
$$

$$y_{11} = \gamma \frac{\left(F_{2s} h_{GD} - F_{2d} h_{GS}\right)\left(g_{1s} - g_{1d}\right) + \left(F_{1d} h_{GS} - F_{1s} h_{GD}\right)\left(g_{2s} - g_{2d}\right)}{B\left(F_{1s} F_{2d} - F_{1d} F_{2s}\right)}$$

$$y_{12} = h_{GD} \frac{F_{2s}\left(g_{1s} - g_{1d}\right) - F_{1s}\left(g_{2s} - g_{2d}\right)}{B\left(F_{1s} F_{2d} - F_{1d} F_{2s}\right)}$$

$$y_{21} = \frac{\left(F_{2s} h_{GS} - F_{2s} h_{GD}\right) g_{1d} + \left(F_{1s} h_{GD} - F_{1d} h_{GS}\right) g_{2d}}{B\left(F_{1s} F_{2d} - F_{1d} F_{2s}\right)}$$

$$y_{22} = \left(h_{GD} + h_{BD}\right) \frac{F_{1s} g_{2d} - F_{2s} g_{1d}}{B\left(F_{1s} F_{2d} - F_{1d} F_{2s}\right)}$$

$$(20)$$

where $\gamma = C'_r/C'$; $F_{1s} = F_1(v_{cs})$; $F_{1d} = F_1(v_{cd})$; $F_{2s} = F_2(v_{cs})$; $F_{2d} = F_2(v_{cd})$; $g_{1s} = g_1(v_{cs})$; $g_{1d} = g_1(v_{cd})$; $g_{2s} = g_2(v_{cs})$; $g_{2d} = g_2(v_{cd})$; and:

$$
\begin{aligned}
h_{GX} &= \frac{C'_r}{-\left(-C' \pm k v_{cs}\right)} \\
h_{BX} &= \frac{C_b}{C'_r} h_{GX}
\end{aligned}
\qquad (21)
$$

The positive (negative) sign in (21) applies when $C'_r(V_G - V_{G0} - V_X)) + C'_b(V_B - V_{B0} - V_X) < 0 \ (>0)$, where the subscript $X$ stands for drain ($D$) and source ($S$).

### B. First-order NQS equivalent circuit of a GFET

If second- and higher-order terms in $\omega$ are neglected when computing (20) to get the NQS $y$-parameters of a GFET, a first-order NQS model is obtained with the following admittance parameters:

$$
\begin{aligned}
y_{11} &= j\omega g_{m0} \gamma \frac{\tau_4}{1 + j\omega \tau_1} & y_{21} &= g_{m0} \frac{1 - j\omega \tau_2}{1 + j\omega \tau_1} \\
y_{12} &= -j\omega g_{m0} \frac{\tau_2}{1 + j\omega \tau_1} & y_{22} &= g_{d0} \frac{1 + j\omega \tau_3}{1 + j\omega \tau_1}
\end{aligned}
\qquad (22)
$$

where:

$$
\begin{aligned}
g_{m0} &= -\mu \frac{W}{L} \frac{k^2}{2C} \left( h_{GD} \left| v_{cd} \right|^3 - h_{GS} \left| v_{cs} \right|^3 \right) \\
g_{d0} &= \mu \frac{W}{L} \frac{k^2}{2C} \left( h_{GD} + h_{BD} \right) \left| v_{cd} \right|^3 \\
\tau_1 &= -\frac{D'}{12} \frac{v_{cd}^4 v_{cs}^4}{v_{cd}^2 + v_{cs}^2} \\
\tau_2 &= \frac{D'}{24} \frac{\left( v_{cd}^6 - 3 v_{cd}^2 v_{cs}^4 + 2 v_{cs}^6 \right)}{h_{GD} v_{cd}^3 - h_{GS} v_{cs}^3} h_{GD} v_{cd}^3 \\
\tau_3 &= -\frac{D'}{24} \left( v_{cd}^6 - 3 v_{cd}^2 v_{cs}^4 + 2 v_{cs}^6 \right) \\
\tau_4 &= \frac{D'}{24} \frac{\left( v_{cd}^2 - v_{cs}^2 \right)^2}{\left( h_{GD} v_{cd}^3 - h_{GS} v_{cs}^3 \right)} \left[ h_{GD} v_{cd}^3 \left( v_{cd}^2 + 2 v_{cs}^2 \right) + h_{GS} v_{cs}^3 \left( 2 v_{cd}^2 + 2 v_{cs}^2 \right) \right]
\end{aligned}
\qquad (23)
$$

where $D' = D/\omega$.

We can go from the equivalent circuit depicted in Fig. 2A to the one shown in Fig. 2B by adopting the following relations:

$$
\begin{aligned}
y_1 &= y_{11} + y_{12} & y_2 &= -y_{12} \\
y_m &= y_{21} - y_{12} & y_0 &= y_{22} + y_{12}
\end{aligned}
\qquad (24)
$$

Therefore, the admittances in Fig. 2B can be expressed as:

$$
\begin{aligned}
y_1 &= j\omega g_{m0} \frac{\left( \gamma \tau_4 - \tau_2 \right)}{1 + j\omega \tau_1} & y_2 &= j\omega g_{m0} \frac{\tau_2}{1 + j\omega \tau_1} \\
y_m &= \frac{g_{m0}}{1 + j\omega \tau_1} & y_0 &= g_{d0} \frac{1 + j\omega \tau_3 \left( 1 - \gamma \right)}{1 + j\omega \tau_1}
\end{aligned}
\qquad (25)
$$

Equation (25) yields an equivalent circuit in which the gate-source and gate-drain admittances, $y_1$ and $y_2$ respectively, are simple $RC$ networks, and the drain-source is formed by the parallel combination of a frequency dependent current source $y_m v_{gs}$ and the output admittance which is a simple lossy $LC$ network. Fig. 2C shows the first-order NQS equivalent circuit of a GFET based on lumped elements, which are quantitatively described by the following formulas:

$$
\begin{aligned}
C_1 &= g_{m0} \left( \gamma \tau_4 - \tau_2 \right) & C_3 &= g_{d0} \tau_3 \left( 1 - \gamma \right) \\
r_1 &= \frac{\tau_1}{C_1} & r_3 &= \frac{\tau_1}{C_3} \\
C_2 &= g_{m0} \tau_2 & r_0 &= \frac{1}{g_{d0}} \\
r_2 &= \frac{\tau_1}{C_2} & L_0 &= \frac{\tau_1}{g_{d0}}
\end{aligned}
\qquad (26)
$$

Note that an $n$-order NQS model of a GFET can be derived by taking higher order terms in $\omega$ in both $g_1(v)$ and $g_2(v)$ in (17) (see SM Subsections C1 and C2 to check the detailed derivation of the zero- and first-order NQS models, respectively).

## III. RESULTS AND DISCUSSION

As a convenient testbed to assess the NQS model, we have selected the prototype GFET described in Table I. The transistor is double-gated with 10-nm $Al_2O_3$ and 300-nm $SiO_2$ dielectrics as top and back gates, respectively. First, Fig. 3A presents the DC transfer characteristics for $V_D = 1V$, $V_S = V_B = 0V$ and Fig. 3B shows the gate bias dependence of the chemical potentials $v_{cs}$ and $v_{cd}$, computed according to (8). The bias window in which the NQS model previously presented cannot be used is indicated by a shaded region, corresponding to biases where the graphene channel is bipolar and/or the consideration $v^2 > \alpha = 1.2 \cdot 10^{-2} eV^2$, i.e., $|v| > 110 meV$, is not satisfied. We have considered that the device operates at room temperature ($T = 300K$) and the inhomogeneity of the electrostatic potential due to the presence of electron-hole puddles is 100 meV causing a residual density of $\sigma_{pud}/q = 6.9 \cdot 10^{15} \text{m}^{-2}$. The resulting gate bias region in which the NQS model cannot be applied is approximately from -0.3V to 1.3V. Out of this bias window, the case B applies for the specific device arrangement considered here.

| $L$ | 1 μm | $V_{G0}$ | 0 V |
|---|---|---|---|
| $W$ | 10 μm | $V_{B0}$ | 0 V |
| $C'_t$ | 7.97 mF/m² | $\mu_0$ | 0.2 m²/Vs |
| $C'_b$ | 115.1 μF/m² | $T$ | 300 K |

Fig. 3C shows the chemical potential along the normalized channel length at $V_G = 1.5$V. In doing so, assuming that the DC current $I_{DS}$ is the same everywhere along the channel, the chemical potential along the channel can be described as follows:

$$v = V_c(x) = \text{Sign}[v_{cd}]\left(v_{cd}^4 - \frac{8C\,I_{DS}\text{Sign}[v_{cd}]}{W\mu k}(L-x)\right) \quad (27)$$

Both $v_{cs} = V_c(0)$ and $v_{cd} = V_c(L)$ have same sign, meaning that the channel is occupied by the same kind of carrier, in this case, electrons.

Now, we superimpose a 4mV sinusoidal time-varying signal at the drain terminal. Fig. 4A shows the absolute value of $i_s$ and $i_d$ for different frequencies, which have been computed according to (12) by truncating the modified Bessel function of the first kind to $n = 10$ (see SM Subsection B3). Table II describes the QS small-signal parameters describing the device at the considered bias point. For convenience, the equivalent circuit of the complete QS small-signal model of GFETs [26] has been depicted in Fig. 2D. The parameters have been calculated according to the QS model presented by the authors in [7],[8]. The cut-off frequency, $f_T = 38.8$GHz, has also been marked in Fig. 4A. It has been analytically calculated using the derivation carried out by the authors in [26], with the small-signal parameters given in Table II and considering extrinsic resistances of $R_sW = R_dW = 100\Omega\mu$m and $R_g = 10\Omega$. According to Fig. 4A, the modulus of $i_s$ and $i_d$ are the same for frequencies lower than ~$f_T/4$, which agrees with the quasi-static assumption, similar to what happens in silicon based FETs [19]. For frequencies higher than $f_T/4$, $|i_s|$ and $|i_d|$ depart from the QS value presenting different values, which indicates that the channel charge in the graphene layer cannot follow the voltage variations for such frequencies.

Fig. 4B shows the modulus of the AC current, $|i|$, along the normalized channel length for different frequencies. In doing so, (6) is computed by using the result of evaluating (27) (see Fig. 3C). The solid orange line depicted in Fig. 4B corresponds to the frequency of $f_T/4$, which has been chosen to delimit the QS and NQS operating zones [19]. For frequencies lower than $f_T/4$, e.g., $f_T/10$, the AC current is approximately the same along the channel length, which agrees with the QS approximation. However, for frequencies higher than $f_T/4$, e.g., $f_T$ or $2f_T$, the carriers do not have enough time to move from drain to source in a signal period, resulting in a departure of $i$ from the QS value, which is in accordance with the behavior observed in Fig. 4A. Therefore, an NQS model of GFETs for the performance prediction and/or circuit simulations for such frequencies is mandatory.

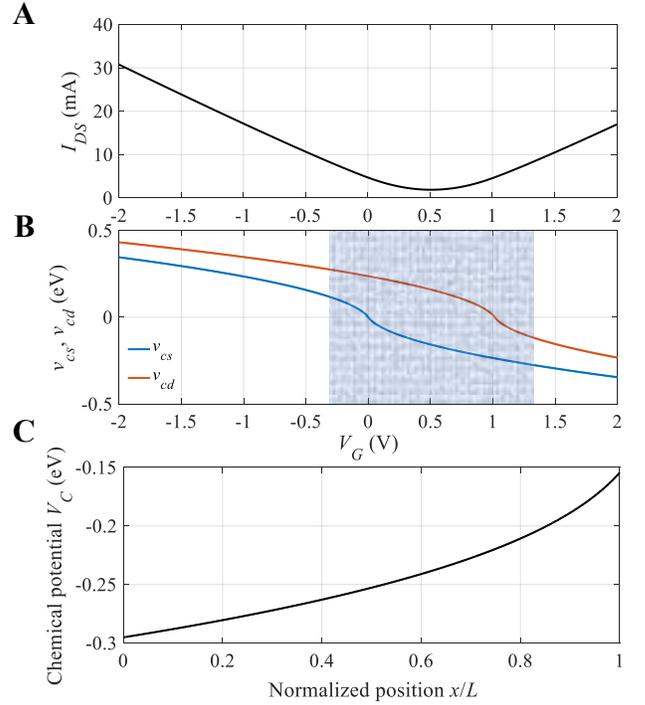

Fig. 3. (A) Transfer characteristics of the device described in Table 1 biased at $V_D = 1$V. (B) Gate bias dependence of the source ($v_{cs}$) and drain ($v_{cd}$) chemical potentials at the same drain voltage. The shaded region indicates the bias region where the NQS model developed in this work cannot be applied. (C) Chemical potential along the normalized position in the channel. The bias point is $V_G = 1.5$V, $V_D = 1$V, $V_B = V_S = 0$V.

| $g_m$ | 13.70 mS | $g_{ds}$ | 7.92 mS |
|---|---|---|---|
| $C_{gs}$ | 43.35 fF | $C_{dg}$ | 31.33 fF |
| $C_{gd}$ | 20.99 fF | $C_{sd}$ | -5.98 fF |

### A. First-order NQS model of the GFET under test

Table III gives the parameters describing the first-order NQS model of the GFET described in Table I for the chosen bias point ($V_G = 1.5$V, $V_D = 1$V, $V_B = V_S = 0$V). They have been calculated by computing (23) and (26). It must be highlighted the similarity shown between the values of $g_m$ and $g_{m0}$; $C_1$ and $C_{gs}$; and $C_2$ and $C_{gd}$, respectively. All first-order NQS small-signal parameters depend on the DC chemical potentials at the drain and source edges shown in Fig. 3B. This way Fig. 5 shows the DC gate bias dependence of such parameters. Therefore, the first-order NQS equivalent circuit of GFETs presented in Fig. 2C can be straightforwardly implemented in Verilog-A and included in any standard circuit simulator forming, to the best of our knowledge, the first compact NQS small-signal model of a GFET.

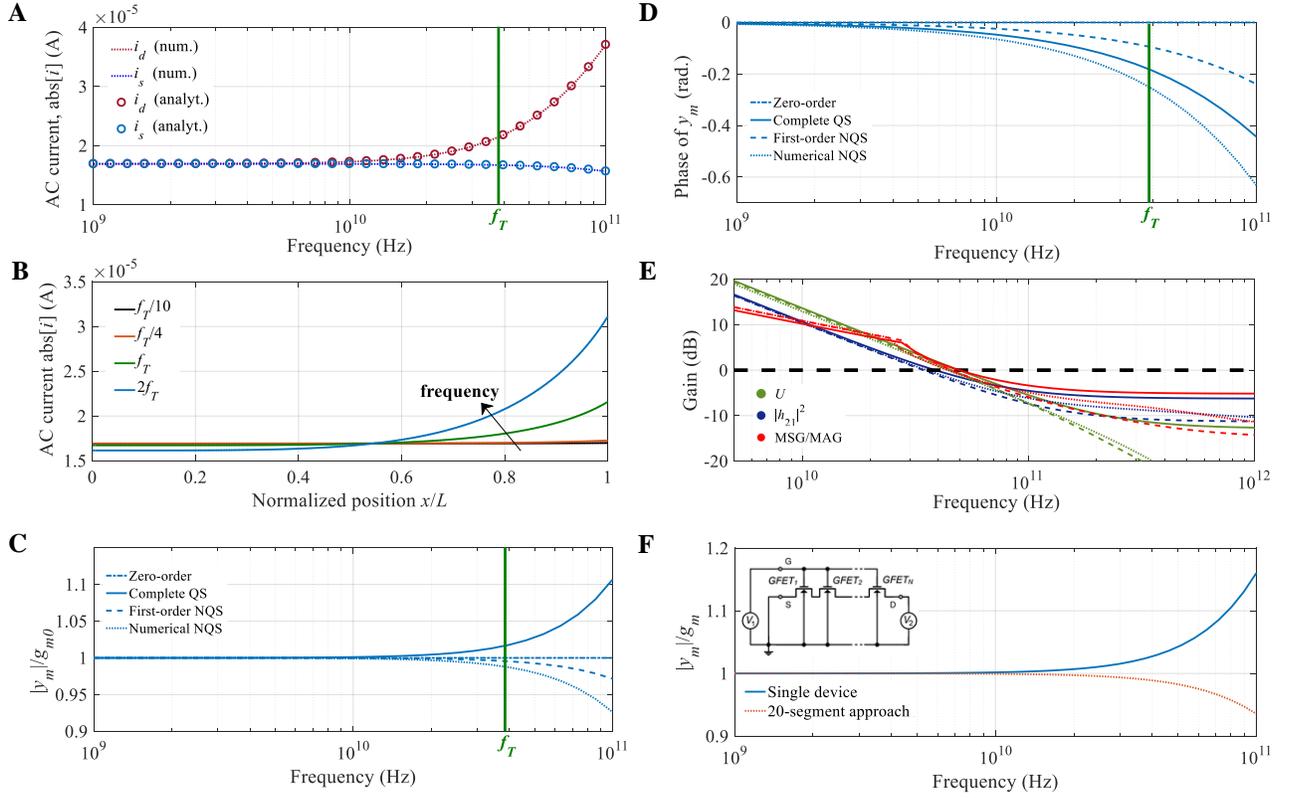

Fig. 4. (A) Modulus of the AC current at the drain ($i_d$) and source ($i_s$) edges computed in an analytical (symbols) and numerical (dotted line) way. The former computed by truncating the modified Bessel function of the first kind to $n = 10$. For the numerical calculation, the function "besseli" of Matlab© software is used. (B) Modulus of the AC current along the normalized position in the channel for different frequencies. (C) Normalized magnitude and (D) phase of $y_m$ versus frequency at $V_G = 1.5$V, $V_D = 1$V, $V_B = V_S = 0$V. Four kind of models are considered: zero-order model (dashdotted lines); the complete QS model described by the parameters given in Table 2 (solid blue lines); the first-order NQS model described by the parameters given in Table 3 (dashed lines); and the numerical NQS model (dotted lines). (E) Small-signal current gain ($h_{21}$), unilateral power gain ($U$, Mason's invariant [30]) and maximum stable gain / maximum available gain (MSG/MAG) versus frequency of the GFET under test predicted by the QS (solid lines), first-order NQS (dashed lines) and numerical NQS (dotted lines) models. The RF figures of merit $f_T$ and $f_{max}$ are gotten when such gains are reduced to unity (0 dB) (see Appendix B for more information). (F) Normalized magnitude of $y_m$ versus frequency under the operating bias point $V_G = 1.5$V, $V_D = 1$V, $V_B = V_S = 0$V for a single 1µm-length GFET compared against a two-port configuration of a cascade of 20 GFETs, 50-nm-length each one, connected in series. (inset) Schematics of the multi-segment approach applied to a GFET.

## TABLE III
### SMALL-SIGNAL ELEMENTS OF THE FIRST-ORDER NQS EQUIVALENT CIRCUIT (FIG. 2C) DESCRIBING THE GFET UNDER TEST

| | | | |
|---|---|---|---|
| $g_{m0}$ | 12.60 mS | $g_{ds0}$ | 4.23 mS |
| $C_1$ | 44.18 fF | $L_0$ | 91.61 pH |
| $r_1$ | 8.78 Ω | $r_0$ | 236.25 Ω |
| $C_2$ | 16.47 fF | $\tau_1$ | 387.76 fs |
| $r_2$ | 23.53 Ω | $\tau_2$ | 1.31 ps |
| $C_3$ | 0.24 fF | $\tau_3$ | 3.95 ps |
| $r_3$ | 1.63 kΩ | $\tau_4$ | 4.88 ps |

### B. Comparison among high-frequency models

A benchmarking of the NQS model against the QS model for the GFET under test is carried out. In doing so, Figs. 4C and 4D show the normalized magnitude and phase of the admittance $y_m$ given by (25), respectively. It is observed that going from the zero-order model to the first-order NQS model produces a drastic improvement in the region of validity. The region of validity for the complete QS model is limited by the fact that, at high frequencies, the error in the magnitude becomes severe. This is because $y_m$ contains a left-half-plane zero for this model [26] in contrast to the left-half-plane pole in $y_m$ for the first-order NQS model (see (25)). The upward-going magnitude predicted by the QS model at high frequencies is clearly unrealistic, since it suggests an enhancement in the forward gate-to-drain action, contrary to the expectation that, at high frequencies, control of the gate on the drain current is gradually lost due to the carrier inertia in the graphene channel. However, the phase of $y_m$ is better predicted by the QS model than does the first-order NQS model, so a higher order correction would be needed if the phase of $y_m$ is crucial for the targeted range of frequency and intended application. This discussion as well as results shown in Figs. 4C and 4D are in agreement with the NQS studies carried out for conventional Si MOSFETs [19],[20].

Next, we have compared the frequency dependence of the current and power gains, namely $|h_{21}|^2$, $U$ and MSG/MAG, as predicted by the different models. In Appendix B we have given further details of RF performance calculation in FETs. The result is shown in Fig. 4E. It can be seen from the plot that the differences between both QS and NQS model predictions are not significant below $f_T$ (38.8 GHz), but clearly the QS model overpredicts the gain in the frequency range 100 GHz – 1 THz, highlighting the importance of including the NQS effects.

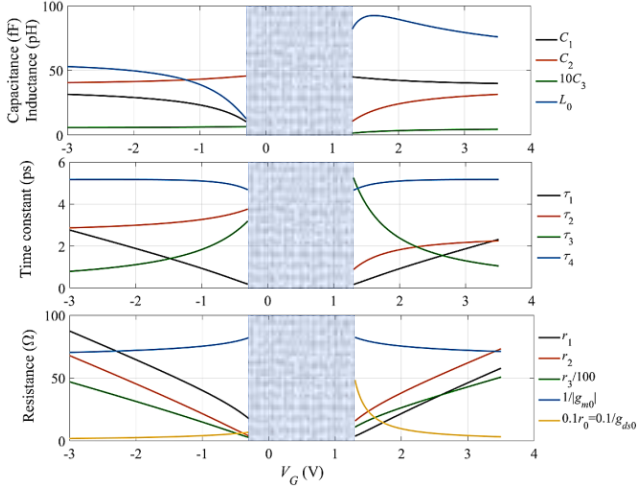

Fig. 5. DC gate bias dependence of the small-signal parameters of the first-order NQS model of GFET. The shaded region indicates the bias region where the NQS model developed in this work cannot be applied.

### C. Multi-segment approach

One way to model a transistor at speeds where the QS model breaks down is to view it as consisting of several sections, each section being short enough to be modeled quasi-statically [19], [27],[28]. To test such an approach we have run some simulations with the circuit simulator Advance Design Systems© by using the intrinsic QS compact model developed by Pasadas and Jiménez [7]. Fig. 4F shows the normalized magnitude of $y_m$ for the intrinsic GFET described in Table I together with the result obtained by a cascade of 20 GFETs in series (connecting the drain of each GFET with the source of the next one and all the device gates are short-circuited), each having a length $L/20 = 50$ nm, and both examined cases at the same bias point previously considered ($V_G = 1.5$V, $V_D = 1$V, $V_B = V_S = 0$V). It is observed that the results agree with the ones shown in Fig. 4C, which demonstrates the consistency of the NQS model presented here.

## IV. CONCLUSION

An NQS model for the GFET has been proposed aiming to capture the delay between the charge change and voltage change when the device is operated at high frequencies. It has been derived following a DC-to-HF methodology that ends up in a general solution for the AC current, which can be numerically computed in terms of the modified Bessel function of the first kind and analytically by truncating the convergent series describing such a Bessel function. An analytical derivation of the zero- and first-order NQS models have been carried out to ultimately obtain an equivalent circuit based on lumped elements, which can be included in a circuit simulator to carry out simulations of arbitrary circuits based on GFETs operated at high-frequencies.

A benchmarking of the first-order and numerical NQS models has been given against the QS model. The frequency region that the NQS model can manage has been also established which slightly depends on the bias point, the desired accuracy, and whether the magnitude or phase is of most interest. According to our simulations, the QS model works up to $f_T/4$, although a first-order NQS model does extend the range of considered frequencies up to $f_T$ accepting an error in $|y_m|/g_{m0}$ < 1% or even $2f_T$ accepting an error < 3.5%. To further validate the NQS model, we have compared its outcome against simulations based on a multi-segment approach where each segment is described by the QS model.

## APPENDIX A

The following notation has been considered for the model formulation:

- Charge per unit area: $Q'(x)$
- Capacitance per unit area: $C'$
- Total charge: $Q = W \int Q'(x) dx$
- Total capacitance: $C = WLC'$
- DC quantities (upper-case symbol with capital subscript): $V_C(x)$
- Large-signal quantities (lower-case symbol with capital subscript): $v_C(x,t)$
- Small-signal quantities (lower-case symbol with lower-case subscript): $v_c(x,t)$
- Time-independent phasor quantities (upper-case symbol with lower-case subscript): $V_c(x,\omega)$

Besides, the following specific notation is also used when the time-varying excitation is considered small and sinusoidal:

- DC channel chemical potentials:
  $$v = V_C(x) \qquad v_{cs} = V_C(0) \qquad v_{cd} = V_C(L)$$
- Time-independent phasors of the chemical potential:
  $$u = V_C(x,\omega) \qquad u_{cs} = V_c(0,\omega) \qquad u_{cd} = V_c(L,\omega)$$
- Time-independent phasors of the current:
  $$i = I_{ds}(x,\omega) \qquad i_s = I_{ds}(0,\omega) \qquad i_d = I_{ds}(L,\omega)$$
- DC current:
  $$I_{DS} = I_{DS}(x)$$

## APPENDIX B

To benchmark an RF technology, it is common to evaluate the cut-off frequency ($f_T$) and the maximum oscillation frequency ($f_{max}$). The computation of such figures of merit can be done from the $y$-parameter matrix. An extrinsic resistance matrix is added to account for the gate resistance ($R_g$) as well as the series combination of the contact and access resistances at the source ($R_s$) and drain ($R_d$) sides which are of upmost importance when dealing with low dimensional FETs [26]. The resulting $y$-parameter matrix reads as follows:

$$\begin{pmatrix} y_{11,x} & y_{12,x} \\ y_{21,x} & y_{22,x} \end{pmatrix} = \left[ \begin{pmatrix} y_{11} & y_{12} \\ y_{21} & y_{22} \end{pmatrix}^{-1} + \begin{pmatrix} R_g + R_s & R_s \\ R_s & R_d + R_s \end{pmatrix} \right]^{-1} \quad (28)$$

The $f_T$ is defined as the frequency at which the magnitude of the small-signal current gain ($h_{21}$) of the transistor is reduced to unity, while the $f_{max}$ is defined as the frequency at which the magnitude of either the unilateral power gain [29] ($U$, or Mason's invariant) or the maximum stable gain / maximum available gain (MSG/MAG) of the transistor is reduced to unity:

$$h_{21} = -\frac{y_{21,s}}{y_{11,s}}$$

$$U = \frac{|y_{12,s} - y_{21,s}|^2}{4\left(\mathrm{Re}\big[y_{11,s}\big]\mathrm{Re}\big[y_{22,s}\big] - \mathrm{Re}\big[y_{12,s}\big]\mathrm{Re}\big[y_{21,s}\big]\right)}$$

$$G_P^{max} = \begin{cases} \mathrm{MSG} = \left|\dfrac{y_{21,s}}{y_{12,s}}\right| & -1 < K < 1 \\[2ex] \mathrm{MAG} = \left|\dfrac{y_{21,s}}{y_{12,s}}\right|\left(K - \sqrt{K^2-1}\right) & K \geq 1;\ |\Delta| < 1 \end{cases} \tag{29}$$

where $K$ and $\Delta$ are the factors used for the evaluation of the stability and are computed as follows:

$$K = \frac{2\mathrm{Re}\big[y_{11,s}\big]\mathrm{Re}\big[y_{22,s}\big] - \mathrm{Re}\big[y_{12,s}y_{21,s}\big]}{|y_{12,s}y_{21,s}|}$$

$$\Delta = \frac{(Y_0 - y_{11,s})(Y_0 - y_{22,s}) - y_{12,s}y_{21,s}}{(Y_0 + y_{11,s})(Y_0 + y_{22,s}) - y_{12,s}y_{21,s}} \tag{30}$$

where $Y_0 = 20\text{mS}$ is the characteristic admittance.

## Acknowledgement

The authors would also like to thank A. Toral-Lopez for the design of Figure 1A.

**Supplementary Material for:**

# Non-quasi-static effects in graphene field-effect transistor under high frequency operation

Francisco Pasadas,*[a] and David Jiménez [a]

[a] Departament d'Enginyeria Electrònica, Escola d'Enginyeria, Universitat Autònoma de Barcelona, 08193 Bellaterra, Spain.
* francisco.pasadas@uab.es

## Notation

The following notation has been considered for the model formulation:

- Charge per unit area: $Q'(x)$
- Capacitance per unit area: $C'$
- Total charge: $Q = W \int_0^L Q'(x) dx$
- Total capacitance: $C = WLC'$
- DC quantities (upper-case symbol with capital subscript): $V_C(x)$
- Large-signal quantities (lower-case symbol with capital subscript): $v_C(x,t)$
- Small-signal quantities (lower-case symbol with lower-case subscript): $v_c(x,t)$
- Time-independent phasor quantities (upper-case symbol with lower-case subscript): $V_c(x,\omega)$

Besides, the following specific notation is also used when the time-varying excitation is considered small and sinusoidal:

- $v = V_C(x)$      $v_{cs} = V_C(0)$      $v_{cd} = V_C(L)$
- $u = V_c(x,\omega)$      $u_{cs} = V_c(0,\omega)$      $u_{cd} = V_c(L,\omega)$
- $i = I_{ds}(x,\omega)$      $i_s = I_{ds}(0,\omega)$      $i_d = I_{ds}(L,\omega)$
- $I_{DS} = I_{DS}(x)$

## A. Detailed derivation of the non-quasi-static model of a GFET

A non-quasi-static (NQS) model, applicable to the high-frequency operation of GFET, is analytically derived by following a similar methodology as the one used for silicon based MOSFETs [1], making the appropriate changes to capture the specific physics of graphene. For such a purpose, it is convenient to study the GFET under different types of excitation, which are covered in the different subsections A1-A4:

A1.      GFET under DC (bias) excitation.
A2.      GFET under time-varying excitation.
A3.      GFET under small-signal time-varying excitation: a special case of A2 where the time-varying excitations are small signals.
A4.      GFET under sinusoidal small-signal time-varying excitation: a special case of A3 where the small signals are sinusoidal.

Before going through the different subsections, we need to define the notation used in this work:

- Charge per unit area: $Q'(x)$
- Capacitance per unit area: $C'$
- Total charge: $Q = W \int_0^L Q'(x) dx$
- Total capacitance: $C = WLC'$
- DC quantities (upper-case symbol with capital subscript): $V_C(x)$
- Large-signal quantities (lower-case symbol with capital subscript): $v_C(x,t)$
- Small-signal quantities (lower-case symbol with lower-case subscript): $v_c(x,t)$
- Time-independent phasor quantities (upper-case symbol with lower-case subscript): $V_c(x,\omega)$

## A1.    GFET under __DC__ (bias) excitation

In this subsection, a four-terminal graphene based device, like the one shown in Figure 1 of the main manuscript, is considered under DC bias excitations.

### Electrostatics

The GFET chemical potential at a channel position $x$, $V_C(x)$, can be read as [2]:

$$V_C(x) = \frac{(C_t' + C_b') - \sqrt{(C_t' + C_b')^2 \pm 2k[C_t'(V_G - V_{G0} - V(x)) + C_b'(V_B - V_{B0} - V(x))]}}{\pm k} \quad \text{(A1.1)}$$

where the positive sign applies when $C_t'(V_G - V_{G0} - V(x)) + C_b'(V_B - V_{B0} - V(x)) < 0$, and the negative sign otherwise. $C_t' = \varepsilon_0 \varepsilon_t / t_t$ ($C_b' = \varepsilon_0 \varepsilon_b / t_b$) is the top (back) oxide capacitance per unit area where $\varepsilon_t$ ($\varepsilon_b$) and $t_t$ ($t_b$) are the top (back) gate oxide relative permittivity and thickness, respectively; and $V_G - V_{G0}$ ($V_B - V_{B0}$) is the top (back) gate voltage overdrive. These quantities comprise any work-function differences between the gates and the graphene channel and any possible additional charges due to impurities or doping. The energy $-qV(x)$ is the quasi-Fermi level along the channel, which must fulfil the boundary conditions: 1) $V(x = 0) = V_S$ (source bias) at the source end; 2) $V(x = L) = V_D$ (drain bias) at the drain end.

In addition, the quantity $dV/dV_C$ can be written as follows [2]:

$$\frac{dV}{dV_C} = 1 + \frac{k|V_C(x)|}{C_t' + C_b'} \quad \text{(A1.2)}$$

### Drain current assuming a drift-diffusion transport mechanism

The static drain-to-source current can be written as follows:

$$I_{DS}(x) = \mu W Q_T'(x)\frac{dV(x)}{dx} = \mu W \left[\frac{k}{2}(V_C^2(x) + \alpha)\right]\left[1 + \frac{k|V_C(x)|}{C_t' + C_b'}\right]\frac{dV_C(x)}{dx} \quad \text{(A1.3)}$$

where the transport charge per unit area, $Q_T'(x)$, is defined in the main manuscript (see Equation (1)).  Then, considering that the current is the same everywhere along the channel:

$$I_{DS} = \mu \frac{W}{L}\frac{k}{2}\int_{V_{CS}}^{V_{CD}}(V_C^2 + \alpha)\left(1 + \frac{k|V_C|}{C_t' + C_b'}\right)dV_C \quad \text{(A1.4)}$$

where:

$$V_{CS} = V_C(x = 0) = \frac{C' - \sqrt{C'^2 \pm 2k[C_t'(V_G - V_{G0} - V_S) + C_b'(V_B - V_{B0} - V_S)]}}{\pm k}$$

$$V_{CD} = V_C(x = L) = \frac{C' - \sqrt{C'^2 \pm 2k[C_t'(V_G - V_{G0} - V_D) + C_b'(V_B - V_{B0} - V_D)]}}{\pm k} \quad \text{(A1.5)}$$

and $C' = C_t' + C_b'$. The gate and substrate currents under DC excitations are considered $I_G = I_B = 0$.

### Charge per unit area associated to each terminal

The top and back gate charges per unit area are [2]:

$$Q_G'(x) = -\frac{C_t'}{C'}Q_{NET}'(x) + Q_0'$$

$$Q_B'(x) = -\frac{C_b'}{C'}Q_{NET}'(x) - Q_0' \quad \text{(A1.6)}$$

where $Q'_{NET}(x)$ and $Q'_0$ are taken from the electrostatics:

$$Q'_{NET}(x) = -C'_t(V_G - V_{G0} - V(x) + V_C(x)) - C'_b(V_B - V_{B0} - V(x) + V_C(x))$$
$$Q'_0 = \frac{C'_t C'_b}{C'}(V_G - V_{G0} - V_B + V_{B0}) \tag{A1.7}$$

Then, the charge controlled by both the drain and source terminals is computed based on the Ward-Dutton's linear charge partition scheme [3], which guarantees charge conservation:

$$Q'_D(x) = \frac{x}{L} Q'_{NET}(x)$$
$$Q'_S(x) = \left(1 - \frac{x}{L}\right) Q'_{NET}(x) \tag{A1.8}$$

## A2.  GFET under <u>time-varying</u> excitation

In this subsection, the four-terminal graphene based device is considered to be fed with time-varying excitations. Therefore, the time-varying chemical potential can be written as follows:

$$v_C(x,t) = \frac{C' - \sqrt{C'^2 \pm 2k[C'_t(v_G(t) - V_{G0} - v^*(x,t)) + C'_b(v_B(t) - V_{B0} - v^*(x,t))]}}{\pm k} \tag{A2.1}$$

Similarly, the time-varying transport charge per unit area reads as:

$$q'_T(x,t) = \frac{k}{2}(v_C^2(x,t) + \alpha) \tag{A2.2}$$

and, therefore, the time dependent drain-to-source current is:

$$i_{DS}(x,t) = \mu W q'_T(x,t) \left[1 + \frac{k|v_C(x,t)|}{C'}\right] \frac{dv_C(x,t)}{dx} \tag{A2.3}$$

Now, the current is no longer the same everywhere along the channel because fast variations are allowed. Instead, the **charge continuity equation** must be considered:

$$\frac{\partial i_{DS}(x,t)}{\partial x} = W \frac{\partial q'_T(x,t)}{\partial t} = W k v_C(x,t) \frac{\partial v_C(x,t)}{\partial t} \tag{A2.4}$$

As a consequence, the terminal currents now are given by:

$$i_D(t) = i_{DS}(L,t) \qquad i_S(t) = i_{DS}(0,t)$$
$$i_G(t) = \frac{dq_G(t)}{dt} \qquad i_B(t) = \frac{dq_B(t)}{dt} \tag{A2.5}$$

where:

$$q_G(t) = W \int_0^L q'_G(x,t)dx \qquad q'_G(x,t) = -\frac{C'_t}{C'} q'_{NET}(x,t) + Q'_0$$
$$q_B(t) = W \int_0^L q'_B(x,t)dx \qquad q'_B(x,t) = -\frac{C'_b}{C'} q'_{NET}(x,t) - Q'_0 \tag{A2.6}$$

and:

$$q'_{NET}(x,t) = -C'_t(v_G(t) - V_{G0} - v^*(x,t) + v_C(x,t))$$
$$- C'_b(v_B(t) - V_{B0} - v^*(x,t) + v_C(x,t)) \tag{A2.7}$$

### A3. GFET under underlined{small-signal} time-varying excitation

When electronic devices operate in analog and RF applications, their terminals are biased with a DC voltage over which a time-varying signal of small amplitude is superimposed. This way the terminal voltages can be written in the following form:

$$v_G(t) = V_G + v_g(t)$$
$$v_B(t) = V_B + v_b(t)$$
$$v_D(t) = V_D + v_d(t)$$

As a result of the above form of the terminal voltages, other time-varying quantities of interest can be expressed as:

$$q'_G(x,t) = Q'_G(x) + q'_g(x,t)$$
$$v_c(x,t) = V_C(x) + v_c(x,t)$$
$$v^*(x,t) = V(x) + v(x,t)$$
$$i_{DS}(x,t) = I_{DS}(x) + i_{ds}(x,t)$$

In the rest of the subsection, the time-varying quantities are going to be split into the DC and small-signal varying parts.

#### Chemical potential

We first proceed applying the following change of variable: $\begin{cases} z_1 = v_G(t) - v^*(x,t) \\ z_2 = v_B(t) - v^*(x,t) \end{cases}$ to the time-varying chemical potential in (A2.1), then:

$$v_c(z_1, z_2) = \frac{C' - \sqrt{C'^2 \pm 2k[C'_t(z_1 - V_{G0}) + C'_b(z_2 - V_{B0})]}}{\pm k}$$

Considering a first-order approximation of the function $v_c(z_1, z_2)$ of two independent variables $(z_1, z_2)$ around the DC point $[c_1, c_2] = [V_G - V(x), V_B - V(x)]$, we obtain:

$$v_c(z_1, z_2) = f[c_1, c_2] + \frac{\partial v_c}{\partial z_1}[c_1, c_2](z_1 - c_1) + \frac{\partial v_c}{\partial z_2}[c_1, c_2](z_2 - c_2)$$

where the DC component of the chemical potential is $V_C(x) = f[c_1, c_2]$ and the small-signal time-varying part of the chemical potential is $v_c(x,t) = \frac{\partial v_c}{\partial z_1}[c_1, c_2](z_1 - c_1) + \frac{\partial v_c}{\partial z_2}[c_1, c_2](z_2 - c_2)$:

$$\boxed{V_C(x) = \frac{C' - \sqrt{C'^2 \pm 2k[C'_t(V_G - V_{G0} - V(x)) + C'_b(V_B - V_{B0} - V(x))]}}{\pm k}}$$

$$\boxed{v_c(x,t) = -\frac{C'_t(v_g(t) - v(x,t)) + C'_b(v_b(t) - v(x,t))}{\sqrt{C'^2 \pm 2k[C'_t(V_G - V_{G0} - V(x)) + C'_b(V_B - V_{B0} - V(x))]}}}$$

(A3.1)

#### Continuity equation

First, the transport charge per unit area in (A2.2) is split into the DC and small-signal time-varying parts:

$$q'_T(x,t) = Q'_T(x) + q'_t(x,t) = \frac{k}{2}\left[(V_C(x) + v_c(x,t))^2 + \alpha\right]$$

$$\boxed{Q'_T(x) = \frac{k}{2}\left[V_C^2(x) + \alpha\right]}$$

$$\boxed{q'_t(x,t) = kV_C(x)v_c(x,t)}$$

(A3.2)

where the small-signal approximation has been assumed, in which only first-order terms in the AC components are retained [4], [5], implying that $v_c^2(x,t) \ll V_C^2(x) + 2V_C(x)v_c(x,t)$. Therefore, if only first-order terms in the AC components are taken, then, the DC and AC parts of the drain-to-source current equation written in (A2.3) would be read as follows:

$$I_{DS}(x) = \mu W \frac{k}{2}\left(V_C{}^2(x) + \alpha\right)\left(1 + \frac{k|V_C(x)|}{C'}\right)\frac{dV_C(x)}{dx}$$

(A3.3)

$$i_{ds}(x,t) = \mu W \frac{k}{2}\left[v_c(x,t)\left(\frac{k\alpha\,\mathrm{Sign}[V_C(x)]}{C'} + 2V_C(x) + 3V_C(x)\frac{k|V_C(x)|}{C'}\right)\frac{dV_C(x)}{dx} + \left(V_C{}^2(x) + \alpha\right)\left(1 + \frac{k|V_C(x)|}{C'}\right)\frac{\partial v_c(x,t)}{\partial x}\right]$$

where it is considered that $|V_C(x) + v_c(x,t)| = \mathrm{Sign}[V_C(x)]\big(V_C(x) + v_c(x,t)\big)$, implying that the AC component of the chemical potential is small enough so that it does not invert the channel polarity defined by the DC component.

Assuming a small-signal time-varying variation, the continuity equation reads as: $\frac{dI_{DS}(x)}{dx} + \frac{\partial i_{ds}(x,t)}{\partial x} = Wk\big(V_C(x) + v_c(x,t)\big)\left[\frac{dV_C(x)}{dt} + \frac{\partial v_c(x,t)}{\partial t}\right]$. Considering now that $\frac{\partial I_{DS}(x)}{\partial x} = 0$ and $\frac{\partial V_C(x)}{\partial t} = 0$, the continuity equation reduces to:

$$\frac{\partial i_{ds}(x,t)}{\partial x} = Wk V_C(x)\frac{\partial v_c(x,t)}{\partial t}$$

(A3.4)

**Top gate charge per unit area**

The charges per unit area associated to each terminal are also split into the DC and small-signal time varying parts, so we can write:

$$Q'_G(x) + q'_g(x,t) = -\frac{C'_t}{C'}[Q'_{NET}(x) + q'_{net}(x,t)] + Q'_0$$

therefore:

$$Q'_G(x) = -\frac{C'_t}{C'}Q'_{NET}(x) + Q'_0$$

$$q'_g(x,t) = -\frac{C'_t}{C'}q'_{net}(x,t)$$

(A3.5)

where:

$$Q'_{NET}(x) = -C'_t\big(V_G - V_{G0} - V(x) + V_C(x)\big) - C'_b\big(V_B - V_{B0} - V(x) + V_C(x)\big)$$

$$q'_{net}(x,t) = -C'_t\big(v_G(t) - v(x,t) + v_c(x,t)\big) - C'_b\big(v_B(t) - v(x,t) + v_c(x,t)\big)$$

(A3.6)

## A4.   GFET under underline{sinusoidal} small-signal time-varying excitation

One could assume that the small-signal voltages are sinusoids and consider the corresponding small-signal terminal charges and currents in the sinusoidal steady state. We will follow a standard practice and consider a fictitious complex exponential excitations of the form: $v_g(t) = V_g e^{j\omega t}$; $v_b(t) = V_b e^{j\omega t}$; $v_d(t) = V_d e^{j\omega t}$; and $v_s(t) = V_s e^{j\omega t}$. Since the equations relating the various small-signal quantities are linear, each small-signal quantity that results as an effect of the excitations described above in the steady state will also be equal to a complex quantity times $e^{j\omega t}$. In particular, we can write, for example: $v_c(x,t) = V_c(x,\omega)e^{j\omega t}$; or $i_{ds}(t) = I_{ds}(x,\omega)e^{j\omega t}$. As in all cases $e^{j\omega t}$ appears as a common factor on both sides, we can obtain the following differential equation from (A3.3) and (A3.4):

$$\begin{cases} I_{ds}(x,\omega) = \mu W \frac{k}{2}\left[V_c(x,\omega)\left(\frac{k\alpha\,\mathrm{Sign}[V_C(x)]}{C'} + 2V_C(x) + 3V_C(x)\frac{k|V_C(x)|}{C'}\right)\frac{dV_C(x)}{dx} + \left(V_C{}^2(x) + \alpha\right)\left(1 + \frac{k|V_C(x)|}{C'}\right)\frac{\partial V_c(x,\omega)}{\partial x}\right] \\ \frac{\partial I_{ds}(x,\omega)}{\partial x} = j\omega Wk V_C(x)V_c(x,\omega) \end{cases}$$

(A4.1)

where the boundary conditions will be calculated as follows:

$$\begin{cases} \boxed{V_c(0,\omega)=V_{cs}}=-\dfrac{C_t'(V_g-V_s)+C_b'(V_b-V_s)}{\sqrt{C'^2\pm 2k[C_t'(V_G-V_{G0}-V_S)+C_b'(V_B-V_{B0}-V_S)]}} \\[4mm] \boxed{V_c(L,\omega)=V_{cd}}=-\dfrac{C_t'(V_g-V_d)+C_b'(V_b-V_d)}{\sqrt{C'^2\pm 2k[C_t'(V_G-V_{G0}-V_D)+C_b'(V_B-V_{B0}-V_D)]}} \end{cases} \qquad \text{(A4.2)}$$

## B.  Derivation of the non-quasi-static model of four-terminal GFETs

In order to build the NQS model of a GFET, the system of differential equations in (A4.1) must be solved. In doing so, we apply some transformations and approximations in the following to get a second-order differential equation describing the NQS response of the device.

The following notation has been used for the sake of convenience:

- $v = V_C(x)$          $v_{cs} = V_C(0)$          $v_{cd} = V_C(L)$
- $u = V_c(x, \omega)$          $u_{cs} = V_c(0, \omega)$          $u_{cd} = V_c(L, \omega)$
- $i = I_{ds}(x, \omega)$          $i_s = I_{ds}(0, \omega)$          $i_d = I_{ds}(L, \omega)$
- $I_{DS} = I_{DS}(x)$

### B1.  Second-order differential equation describing the AC operation at high-frequencies

The following transformations to the differential equations (A4.1) describing the small-signal sinusoidal operation of the GFET have been applied [4]:

$$I_{DS} = \mu W \frac{k}{2}(v^2 + \alpha)\left(1 + \frac{k|v|}{C'}\right)\frac{dv}{dx} \;\rightarrow\; \frac{dx}{dv} = \frac{\mu W k}{2 I_{DS}}(v^2 + \alpha)\left(1 + \frac{k}{C'}|v|\right)$$

$$i = \mu W \frac{k}{2}\left[u\left(\frac{k\alpha}{C'}\frac{|v|}{v} + 2v + 3v\frac{k|v|}{C'}\right)\frac{dv}{dx} + (v^2 + \alpha)\left(1 + \frac{k|v|}{C'}\right)\frac{du}{dx}\right] \;\rightarrow\; \frac{di}{dx} = j\omega W k v u$$

$$\boxed{\frac{di}{dv} = j\omega \frac{\mu W^2 k^2}{2 I_{DS}} v u(v^2 + \alpha)\left(1 + \frac{k}{C'}|v|\right)} \tag{B1.1}$$

Finally, the following second-order differential equation describes the NQS response of a GFET:

$$\boxed{\frac{d^2 i}{dv^2} - \frac{1}{v}\frac{di}{dv} - j\omega \frac{\mu W^2 k^2}{2 I_{DS}{}^2} v(v^2 + \alpha)\left(1 + \frac{k}{C'}|v|\right) i = 0} \tag{B1.2}$$

However, such an equation cannot be analytically solved. Therefore we rewrite (B1.2) as: $\frac{d^2 i}{dv^2} - \frac{1}{v}\frac{di}{dv} - j\omega \frac{\mu W^2 k^2}{2 I_{DS}{}^2} f[v] i = 0$; where $f[v] = v(v^2 + \alpha)\left(1 + \frac{k}{C'}|v|\right)$. We assume that $v^2 \gg \alpha$, thus, $f[v] \approx v^3\left(1 + \frac{k}{C'}|v|\right)$. The transport charge per unit area $Q_T'(x) = \frac{k}{2}(v^2 + \alpha)$ and the parameter $\alpha = \left(\frac{(\pi k_B T)^2}{3 q^2} + \frac{2}{k}\sigma_{pud}\right)$, are defined in the main manuscript in Equation (1). The term $\alpha \frac{k}{2}$ in $Q_T'(x)$ acts as residual charge density produced by the temperature and possible presence of electron-hole puddles [6]. Therefore, $v^2 \gg \alpha$ holds when the GFET is operated far from the Dirac voltage and/or when the graphene channel has low $\sigma_{pud}$ (which is an indicator of high graphene quality).

In that case, the approximated second-order differential equation can be written as: $\frac{d^2 i}{dv^2} - \frac{1}{v}\frac{di}{dv} - j\omega \frac{\mu W^2 k^2}{2 I_{DS}{}^2} f[v] i = 0$ and $f[v] = v^3\left(1 + \frac{k}{C'}|v|\right)$. Now, we consider two different scenarios allowing further simplification of the differential equation:

- **Case A**: $\frac{k}{C'}|v| \ll 1$, meaning that the parallel combination of top and back geometrical capacitances per unit area is larger than the quantum capacitance of graphene $C' \gg C_q' = k|v|$. Therefore, $f[v] \approx v^3$ and the following second-order differential equation is obtained:

$$\boxed{\frac{d^2 i}{dv^2} - \frac{1}{v}\frac{di}{dv} - j\omega \frac{\mu W^2 k^2}{2 I_{DS}{}^2} v^3 i = 0} \tag{B1.3}$$

- **Case B:** $\frac{k}{C'}|v| \gg 1$, meaning that the parallel combination of top and back geometrical capacitances per unit area is lower than the quantum capacitance of graphene $C' \ll C'_q = k|v|$. Therefore, $f[v] \approx \frac{k}{C'}\text{Sign}[v]v^4$ and the following second-order differential equation is obtained:

$$\boxed{\frac{d^2i}{dv^2} - \frac{1}{v}\frac{di}{dv} - j\omega\frac{\mu W^2 k^3}{2C'I_{DS}{}^2}\text{Sign}[v]v^4 i = 0}$$
(B1.4)

The boundary conditions are:

$$\begin{cases} i[v(x=0) = V_C(0) = v_{cs}] = i_s = I_{ds}(0,\omega) \\ i[v(x=L) = V_C(L) = v_{cd}] = i_d = I_{ds}(L,\omega) \end{cases}$$

$$\begin{cases} \boxed{V_C(0) = v_{cs}} = \dfrac{C' - \sqrt{C'^2 \pm 2k[C'_t(V_G - V_{G0} - V_S) + C'_b(V_B - V_{B0} - V_S)]}}{\pm k} \\[4mm] \boxed{V_C(L) = v_{cd}} = \dfrac{C' - \sqrt{C'^2 \pm 2k[C'_t(V_G - V_{G0} - V_D) + C'_b(V_B - V_{B0} - V_D)]}}{\pm k} \end{cases}$$

$$\begin{cases} \boxed{V_C(0,\omega) = u_{cs}} = -\dfrac{C'_t(V_g - V_s) + C'_b(V_b - V_s)}{\sqrt{C'^2 \pm 2k[C'_t(V_G - V_{G0} - V_S) + C'_b(V_B - V_{B0} - V_S)]}} \\[4mm] \boxed{V_C(L,\omega) = u_{cd}} = -\dfrac{C'_t(V_g - V_d) + C'_b(V_b - V_d)}{\sqrt{C'^2 \pm 2k[C'_t(V_G - V_{G0} - V_D) + C'_b(V_B - V_{B0} - V_D)]}} \end{cases}$$

## B2.  Solution of the second-order differential equation

- **Case A:** $\frac{k}{C'}|v| \ll 1$

The second-order differential equation in (B1.3) can be written as: $\frac{d^2i}{dv^2} - \frac{1}{v}\frac{di}{dv} - jDv^3 i = 0$ where $D = \omega\frac{\mu W^2 k^2}{2I_{DS}{}^2}$. It can be analytically solved, being the general solution:

$$i(v) = v\left[\frac{i_s}{v_{cs}}\frac{n_3(v)p_2 - p_3(v)n_2}{p_2 n_1 - n_2 p_1} + \frac{i_d}{v_{cd}}\frac{p_3(v)n_1 - n_3(v)p_1}{p_2 n_1 - n_2 p_1}\right]$$
(B2.1)

where $p_i$ and $n_i$ stands for:

$$n_1 = I_{-\frac{2}{5}}\left(\frac{2}{5}D_1 v_{cs}^{5/2}\right)$$
$$n_2 = I_{-\frac{2}{5}}\left(\frac{2}{5}D_1 v_{cd}^{5/2}\right)$$
$$n_3(v) = I_{-\frac{2}{5}}\left(\frac{2}{5}D_1 v^{5/2}\right)$$
$$p_1 = I_{\frac{2}{5}}\left(\frac{2}{5}D_1 v_{cs}^{5/2}\right)$$
$$p_2 = I_{\frac{2}{5}}\left(\frac{2}{5}D_1 v_{cd}^{5/2}\right)$$
$$p_3(v) = I_{\frac{2}{5}}\left(\frac{2}{5}D_1 v^{5/2}\right)$$

where $D_1 = \sqrt[4]{-1}\sqrt{D}$ and $I_\lambda(z)$ represents the modified Bessel function of the first kind and real order $\lambda$ of the complex argument $z$.

To obtain the solution for $i(v)$, we have to determine the boundary conditions $i_s$ and $i_d$. To make it possible, we get $\frac{di}{dv}$ from the general solution (B2.1):

$$\frac{di}{dv} = D_1 v^{\frac{5}{2}} \left[ \frac{i_s}{v_{cs}} \frac{p_4(v)p_2 - n_4(v)n_2}{p_2 n_1 - n_2 p_1} + \frac{i_d}{v_{cd}} \frac{n_4(v)n_1 - p_4(v)p_1}{p_2 n_1 - n_2 p_1} \right] \qquad (B2.2)$$

where:

$$n_4(v) = I_{-\frac{3}{5}}\left(\frac{2}{5}D_1 v^{5/2}\right)$$

$$p_4(v) = I_{\frac{3}{5}}\left(\frac{2}{5}D_1 v^{5/2}\right)$$

To derive (B2.2) we have used the following results:

$$\frac{d}{dz}\left[z \cdot I_{-\frac{2}{5}}\left(\frac{2}{5}D_1 z^{5/2}\right)\right] = \frac{d}{dz}[z \cdot n_3(z)] = I_{-\frac{2}{5}}\left(\frac{2}{5}D_1 z^{5/2}\right) + \frac{1}{2}D_1 z^{5/2}\left\{I_{-\frac{7}{5}}\left(\frac{2}{5}D_1 z^{5/2}\right) + I_{\frac{3}{5}}\left(\frac{2}{5}D_1 z^{5/2}\right)\right\}$$

$$\rightarrow \boxed{\frac{d}{dz}\left[z \cdot I_{-\frac{2}{5}}\left(\frac{2}{5}D_1 z^{5/2}\right)\right] = D_1 z^{5/2}I_{\frac{3}{5}}\left(\frac{2}{5}D_1 z^{5/2}\right) = D_1 z^{\frac{5}{2}}p_4(z)}$$

$$\frac{d}{dz}\left[z \cdot I_{\frac{2}{5}}\left(\frac{2}{5}D_1 z^{5/2}\right)\right] = \frac{d}{dz}[z \cdot p_3(z)] = I_{\frac{2}{5}}\left(\frac{2}{5}D_1 z^{5/2}\right) + \frac{1}{2}D_1 z^{5/2}\left\{I_{-\frac{3}{5}}\left(\frac{2}{5}D_1 z^{5/2}\right) + I_{\frac{7}{5}}\left(\frac{2}{5}D_1 z^{5/2}\right)\right\}$$

$$\rightarrow \boxed{\frac{d}{dz}\left[z \cdot I_{\frac{2}{5}}\left(\frac{2}{5}D_1 z^{5/2}\right)\right] = D_1 z^{5/2}I_{-\frac{3}{5}}\left(\frac{2}{5}D_1 z^{5/2}\right) = D_1 z^{\frac{5}{2}}n_4(z)}$$

Now, using (B1.1), we can write: $\frac{di}{dv} = j\omega \frac{\mu W^2 k^2}{2I_0}vu(v^2+y)\left(1 + \frac{k}{c'}|v|\right) \approx j\omega\frac{\mu W^2 k^2}{2I_0}v^3 u$. Making it equal to (B2.2), the following system of equations arise:

$$\begin{cases} \frac{di}{dv}\Big|_{v=v_{cs}} \rightarrow j\omega\frac{\mu W^2 k^2}{2I_{DS}}v_{cs}^3 u_{cs} = D_1 v_{cs}^{\frac{5}{2}}\left[\frac{i_s}{v_{cs}}\frac{p_4(v_{cs})p_2 - n_4(v_{cs})n_2}{p_2 n_1 - n_2 p_1} + \frac{i_d}{v_{cd}}\frac{n_4(v_{cs})n_1 - p_4(v_{cs})p_1}{p_2 n_1 - n_2 p_1}\right] \\ \frac{di}{dv}\Big|_{v=v_{cd}} \rightarrow j\omega\frac{\mu W^2 k^2}{2I_{DS}}v_{cd}^3 u_{cd} = D_1 v_{cd}^{\frac{5}{2}}\left[\frac{i_s}{v_{cs}}\frac{p_4(v_{cd})p_2 - n_4(v_{cd})n_2}{p_2 n_1 - n_2 p_1} + \frac{i_d}{v_{cd}}\frac{n_4(v_{cd})n_1 - p_4(v_{cd})p_1}{p_2 n_1 - n_2 p_1}\right] \end{cases}$$

The solution of such a system of equations is:

$$\begin{cases} i_s = v_{cs}\dfrac{D_2}{D_1}\dfrac{u_{cd}\sqrt{v_{cd}}[n_1 n_{4s} - p_1 p_{4s}] + u_{cs}\sqrt{v_{cs}}[p_1 p_{4d} - n_1 n_{4d}]}{p_{4d}n_{4s} - n_{4d}p_{4s}} \\ i_d = v_{cd}\dfrac{D_2}{D_1}\dfrac{u_{cd}\sqrt{v_{cd}}[n_2 n_{4s} - p_2 p_{4s}] + u_{cs}\sqrt{v_{cs}}[p_2 p_{4d} - n_2 n_{4d}]}{p_{4d}n_{4s} - n_{4d}p_{4s}} \end{cases} \qquad (B2.3)$$

where:

$$D_2 = jDI_{DS}$$
$$n_{4s} = n_4(v_{cs})$$
$$n_{4d} = n_4(v_{cd})$$
$$p_{4s} = p_4(v_{cs})$$
$$p_{4d} = p_4(v_{cd})$$

- **Case B:** $\frac{k}{c'}|v| \gg 1$

The second-order differential equation in (B1.4) can be written as: $\frac{d^2 i}{dv^2} - \frac{1}{v}\frac{di}{dv} - jDv^4 i = 0$ where $D = \omega\frac{\mu W^2 k^3}{2c'I_{DS}^2}\text{Sign}[v]$. It can be analytically solved, being the general solution:

$$i(v) = v\left[\frac{i_s}{v_{cs}}\frac{n_3(v)p_2 - p_3(v)n_2}{p_2 n_1 - n_2 p_1} + \frac{i_d}{v_{cd}}\frac{p_3(v)n_1 - n_3(v)p_1}{p_2 n_1 - n_2 p_1}\right] \qquad (B2.4)$$

where $p_i$ and $n_i$ stands for:

$$n_1 = I_{-\frac{1}{3}}\left(\frac{1}{3}D_1 v_{cs}^3\right)$$

$$n_2 = I_{-\frac{1}{3}}\left(\frac{1}{3}D_{1d} v_{cd}^3\right)$$

$$n_3(v) = I_{-\frac{1}{3}}\left(\frac{1}{3}D_1 v^3\right)$$

$$p_1 = I_{\frac{1}{3}}\left(\frac{1}{3}D_{1s} v_{cs}^3\right)$$

$$p_2 = I_{\frac{1}{3}}\left(\frac{1}{3}D_{1d} v_{cd}^3\right)$$

$$p_3(v) = I_{\frac{1}{3}}\left(\frac{1}{3}D_1 v^3\right)$$

where $D_1 = \sqrt[4]{-1}\sqrt{D}$; $D_{1s} = D_1|_{v=v_{cs}}$; $D_{1d} = D_1|_{v=v_{cd}}$ and $I_\lambda(z)$ represents the modified Bessel function of the first kind and real order $\lambda$ of the complex argument $z$.

To obtain the solution for $i(v)$, we have to determine the boundary conditions $i_s$ and $i_d$. Again, from the general solution (B2.4) we can get $\frac{di}{dv}$ in the following form:

$$\frac{di}{dv} = D_1 v^3 \left[ \frac{i_s}{v_{cs}} \frac{p_4(v)p_2 - n_4(v)n_2}{p_2 n_1 - n_2 p_1} + \frac{i_d}{v_{cd}} \frac{n_4(v)n_1 - p_4(v)p_1}{p_2 n_1 - n_2 p_1} \right] \tag{B2.5}$$

where:

$$n_4(v) = I_{-\frac{2}{3}}\left(\frac{1}{3}D_1 v^3\right)$$

$$p_4(v) = I_{\frac{2}{3}}\left(\frac{1}{3}D_1 v^3\right)$$

To derive (B2.5) we have used the following results:

$$\frac{d}{dz}\left[z \cdot I_{-\frac{1}{3}}\left(\frac{1}{3}D_1 z^3\right)\right] = \frac{d}{dz}[z \cdot n_3(z)] = I_{-\frac{1}{3}}\left(\frac{1}{3}D_1 z^3\right) + \frac{1}{2}D_1 z^3 \left\{ I_{-\frac{4}{3}}\left(\frac{1}{3}D_1 z^3\right) + I_{\frac{2}{3}}\left(\frac{1}{3}D_1 z^3\right) \right\}$$

$$\rightarrow \boxed{\frac{d}{dz}\left[z \cdot I_{-\frac{1}{3}}\left(\frac{1}{3}D_1 z^3\right)\right] = D_1 z^3 I_{\frac{2}{3}}\left(\frac{1}{3}D_1 z^3\right)} = D_1 z^3 p_4(z)$$

$$\frac{d}{dz}\left[z \cdot I_{\frac{1}{3}}\left(\frac{1}{3}D_1 z^3\right)\right] = \frac{d}{dz}[z \cdot p_3(z)] = I_{\frac{1}{3}}\left(\frac{1}{3}D_1 z^3\right) + \frac{1}{2}D_1 z^3 \left\{ I_{-\frac{2}{3}}\left(\frac{1}{3}D_1 z^3\right) + I_{\frac{4}{3}}\left(\frac{1}{3}D_1 z^3\right) \right\}$$

$$\rightarrow \boxed{\frac{d}{dz}\left[z \cdot I_{\frac{1}{3}}\left(\frac{1}{3}D_1 z^3\right)\right] = D_1 z^3 I_{-\frac{2}{3}}\left(\frac{1}{3}D_1 z^3\right)} = D_1 z^3 n_4(z)$$

Now, coming back to (B1.1), we can write: $\frac{di}{dv} = j\omega \frac{\mu W^2 k^2}{2I_0} v u(v^2 + y)\left(1 + \frac{k}{C'}|v|\right) \approx j\omega \frac{\mu W^2 k^3}{2C'I_{DS}}\text{Sign}[v]v^4 u$. Making it equal to (B2.5), the following system of equations arise:

$$\begin{cases} \left.\frac{di}{dv}\right|_{v=v_{cs}} \rightarrow j\omega \frac{\mu W^2 k^3}{2C'I_0}\text{Sign}[v_{cs}]v_{cs}^4 u_{cs} = D_{1s} v_{cs}^3 \left[ \frac{i_s}{v_{cs}} \frac{p_4(v_{cs})p_2 - n_4(v_{cs})n_2}{p_2 n_1 - n_2 p_1} + \frac{i_d}{v_{cd}} \frac{n_4(v_{cs})n_1 - p_4(v_{cs})p_1}{p_2 n_1 - n_2 p_1} \right] \\ \left.\frac{di}{dv}\right|_{v=v_{cd}} \rightarrow j\omega \frac{\mu W^2 k^3}{2C'I_0}\text{Sign}[v_{cd}]v_{cd}^4 u_{cd} = D_{1d} v_{cd}^3 \left[ \frac{i_s}{v_{cs}} \frac{p_4(v_{cd})p_2 - n_4(v_{cd})n_2}{p_2 n_1 - n_2 p_1} + \frac{i_d}{v_{cd}} \frac{n_4(v_{cd})n_1 - p_4(v_{cd})p_1}{p_2 n_1 - n_2 p_1} \right] \end{cases}$$

The solution of such a system of equations is:

$$\begin{cases} i_s = v_{cs} \dfrac{D_{1s}D_{2d}u_{cd}v_{cd}[n_1 n_{4s} - p_1 p_{4s}] + D_{1d}D_{2s}u_{cs}v_{cs}[p_1 p_{4d} - n_1 n_{4d}]}{D_{1s}D_{1d}[p_{4d}n_{4s} - n_{4d}p_{4s}]} \\ i_d = v_{cd} \dfrac{D_{1s}D_{2d}u_{cd}v_{cd}[n_2 n_{4s} - p_2 p_{4s}] + D_{1d}D_{2s}u_{cs}v_{cs}[p_2 p_{4d} - n_2 n_{4d}]}{D_{1s}D_{1d}[p_{4d}n_{4s} - n_{4d}p_{4s}]} \end{cases} \tag{B2.6}$$

where:

$$D_2 = jDI_{DS}$$
$$D_{2s} = D_2|_{v=v_{cs}}$$

$$D_{2d} = D_2|_{v=v_{cd}}$$
$$n_{4s} = n_4(v_{cs})$$
$$n_{4d} = n_4(v_{cd})$$
$$p_{4s} = p_4(v_{cs})$$
$$p_{4d} = p_4(v_{cd})$$

### B3.   Analytic calculation of the modified Bessel function of the first kind

The general solution of $i(v)$ in both cases, namely (B2.1) and (B2.4), as well as the boundary conditions $i_s$ and $i_d$, given by (B2.3) and (B2.6), respectively, are based on the modified Bessel function of the first kind, $I_\lambda(z)$, where $\lambda$ is the real order and $z$ is the complex argument:

$$I_\lambda(z) = \sum_{k=0}^{\infty} \frac{1}{\Gamma(k + \nu + 1)k!} \left(\frac{z}{2}\right)^{2k+\nu} \tag{B3.1}$$

Such a series is convergent everywhere in the complex $z$-plane.

To build a compact NQS model we propose to truncate the series (B3.1) at a finite positive integer, $n$, which implies accepting a trade-off between accuracy and relative error, namely $I_\nu(z, n) = \sum_{k=0}^{n} \frac{1}{\Gamma(k+\nu+1)k!}\left(\frac{z}{2}\right)^{2k+\nu}$. For example, considering the device described in the main manuscript, the operating bias point chosen, and the frequencies given in Figure 4A of the main manuscript, the maximum relative error between the numerical calculation of the AC currents and the analytic calculation by truncating the convergent series is $\epsilon < 3 \cdot 10^{-5}$ for $n = 5$ and $\epsilon < 7 \cdot 10^{-13}$ for $n = 10$. The relative error has been defined as $\epsilon = \left|\frac{i_{num} - i_{ana}}{i_{num}}\right|$.

On the other hand, to analytically calculate the truncated Gamma function we have used the Lanczos Approximation [7], [8], which is an efficient algorithm for computing the Gamma function for any real and complex argument with a non-negative real part, to a high level of accuracy. The error is smaller than $2 \cdot 10^{-10}$ for any complex $z$ for which $\text{Re}[z] > 0$:

$$\Gamma(z) = \left[\frac{\sqrt{2\pi}}{z}\left(q_0 + \sum_{m=1}^{6} \frac{q_m}{z+m}\right)\right](z + 5.5)^{z+0.5} e^{-(z+5.5)} \tag{B3.2}$$
$$\text{Re}[z] > 0$$

where

$$q_0 = 1.000000000190015$$
$$q_1 = 76.18009172947146$$
$$q_2 = -86.50532032941677$$
$$q_3 = 24.01409824083091$$
$$q_4 = -1.231739572450155$$
$$q_5 = 1.208650973866179 \cdot 10^{-3}$$
$$q_6 = -5.395239384953 \cdot 10^{-6}$$

As we have to calculate $\Gamma(k + \lambda + 1)$, where $k = 0,1,2 \dots n$ and $\lambda = \pm\frac{1}{3}, \pm\frac{1}{2}, \pm\frac{2}{3}, \pm\frac{2}{5}, \pm\frac{3}{5}$, it is guaranteed that the argument of the Gamma function always satisfies $\text{Re}[z] > 0$, so we can use (B3.2) to accurately approximate such function.

### B4.   Compact NQS model of a four-terminal GFET

Figure S1 depicts the schematics of a four-terminal FET operating under a small-signal regime showing the DC and AC voltages and currents. From (B2.3) or (B2.6), depending on the case, the four AC terminal currents are determined assuming that the time-varying signal is sinusoidal:

- AC drain current: $i_d(t) \to I_{ds}(L, \omega) = v_{cd} \dfrac{D_{1s}D_{2d}u_{cd}v_{cd}[n_2n_{4s}-p_2p_{4s}]+D_{1d}D_{2s}u_{cs}v_{cs}[p_2p_{4d}-n_2n_{4d}]}{D_{1s}D_{1d}[p_{4d}n_{4s}-n_{4d}p_{4s}]}$ and
  $i_d(t) = \dfrac{dq_d(t)}{dt} \to I_d(\omega) = j\omega Q_d(\omega)$

- AC source current: $i_s(t) \to -I_{ds}(0, \omega) = -v_{cs} \dfrac{D_{1s}D_{2d}u_{cd}v_{cd}[n_1n_{4s}-p_1p_{4s}]+D_{1d}D_{2s}u_{cs}v_{cs}[p_1p_{4d}-n_1n_{4d}]}{D_{1s}D_{1d}[p_{4d}n_{4s}-n_{4d}p_{4s}]}$
  and $i_s = \dfrac{dq_s(t)}{dt} \to I_s(\omega) = j\omega Q_s(\omega)$

- AC top-gate current: $i_g(t) = \dfrac{dq_g(t)}{dt} \to I_g(\omega) = j\omega Q_g(\omega)$

- AC back-gate current: $i_b(t) = \dfrac{dq_b(t)}{dt} \to I_b(\omega) = j\omega Q_b(\omega)$

where:

$$q'_d(x,t) = \frac{x}{L}q'_{net}(x,t) \qquad\qquad q_d(t) = W\int_0^L q'_d(x,t)dx$$

$$q'_s(x,t) = \left(1-\frac{x}{L}\right)q'_{net}(x,t) \qquad\qquad q_s(t) = W\int_0^L q'_s(x,t)dx$$

$$q'_g(x,t) = -\frac{C'_t}{C'}q'_{net}(x,t) \qquad\qquad q_g(t) = W\int_0^L q'_g(x,t)dx = -\frac{WC'_t}{C'}\int_0^L q'_{net}(x,t)dx$$

$$q'_b(x,t) = -\frac{C'_b}{C'}q'_{net}(x,t) \qquad\qquad q_b(t) = W\int_0^L q'_b(x,t)dx = -\frac{WC'_b}{C'}\int_0^L q'_{net}(x,t)dx$$

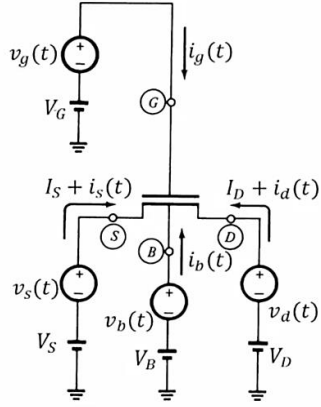

**Figure S1.-** Schematics of a four-terminal FET operating under small-signal regime showing the terminal DC and AC voltages and currents.

**Charge conservation:** $\boxed{q'_g(x,t) + q'_b(x,t) + q'_d(x,t) + q'_s(x,t) = 0}$

In order to calculate $i_g(t)$ and $i_b(t)$, we need to calculate $q_{net}(t) = W\int_0^L q'_{net}(x,t)dx$, which can be obtained from $i_d(t)$ and $i_s(t)$ as follows:

$$i_d(t) + i_s(t) \to j\omega(Q_d(\omega) + Q_s(\omega)) = j\omega W\left(\int_0^L \frac{x}{L}Q'_{net}(x,\omega)dx + \int_0^L\left(1-\frac{x}{L}\right)Q'_{net}(x,\omega)dx\right)$$

$$= j\omega W\int_0^L Q'_{net}(x,\omega)dx \to \boxed{\int_0^L Q'_{net}(x,\omega)dx = \frac{I_d(\omega)+I_s(\omega)}{j\omega W} = \frac{I_{ds}(L,\omega)-I_{ds}(0,\omega)}{j\omega W} = \frac{(i_d-i_s)}{j\omega W}}$$

Therefore:

$$j\omega Q_g(\omega) = -\frac{j\omega WC'_t}{C'}\int_0^L Q'_{net}(x,\omega)dx \to \boxed{i_g = -\frac{C'_t}{C'}(i_d-i_s)}$$

$$j\omega Q_b(\omega) = -\frac{j\omega WC'_b}{C'}\int_0^L Q'_{net}(x,\omega)dx \to \boxed{i_b = -\frac{C'_b}{C'}(i_d-i_s)}$$

(B4.1)

### C. The short-circuit admittance parameters

The short-circuit admittance parameters are gotten by relating the terminal current phasors to the terminal voltage phasors. They can be easily calculated using the definition, taking the source as the reference [1]: $y_{jk} = \frac{i_j}{u_{ks}}\Big|_{u_{ls}=0;\; l\neq k}$, where the subscripts $j$, $k$, and $l$ stand for drain ($d$), top gate ($g$), and back gate ($b$). In this work we have considered that the back gate terminal is also AC short-circuited to the source, thus, the GFET can be seen as a two-port network in common-source configuration (Figure S2), with the following four independent admittance parameters:

$$y_{11} = \frac{i_g}{u_{gs}}\Big|_{u_{ds}=0}$$

$$y_{12} = \frac{i_g}{u_{ds}}\Big|_{u_{gs}=0}$$

$$y_{21} = \frac{i_d}{u_{gs}}\Big|_{u_{ds}=0}$$

$$y_{22} = \frac{i_d}{u_{ds}}\Big|_{u_{gs}=0}$$

where ports 1 and 2 refer to the gate-source and drain-source ports, respectively.

- **Case A:** $\frac{k}{C'}|v| \ll 1$

Equation (B2.1) can be written in the following way

$$i(v) = v\left[\left(\frac{i_d n_1 v_{cs} - i_s n_2 v_{cd}}{v_{cs} v_{cd}(p_2 n_1 - n_2 p_1)}\right)p_3(v) + \left(\frac{i_s p_2 v_{cd} - i_d p_1 v_{cs}}{v_{cs} v_{cd}(p_2 n_1 - n_2 p_1)}\right)n_3(v)\right] = a_1 f_1(v) + a_2 f_2(v)$$

where: $a_1 = \frac{i_d n_1 v_{cs} - i_s n_2 v_{cd}}{v_{cs} v_{cd}(p_2 n_1 - n_2 p_1)}$; $a_2 = \frac{i_s p_2 v_{cd} - i_d p_1 v_{cs}}{v_{cs} v_{cd}(p_2 n_1 - n_2 p_1)}$; $f_1(v) = v \cdot p_3(v)$; and $f_2(v) = v \cdot n_3(v)$.

and:

$$\boxed{p_3(v) = I_{\frac{2}{5}}\left(\frac{2}{5}D_1 v^{5/2}\right) = \sum_{m=0}^{\infty} \frac{1}{\Gamma\left(m+\frac{7}{5}\right)m!}\left(\frac{D_1 v^{5/2}}{5}\right)^{2m+\frac{2}{5}}} \qquad (C.1)$$

- $m=0$    $p_3(v) = \left(\frac{D_1}{5}\right)^{2/5}\frac{1}{\Gamma\left(\frac{2}{5}\right)}v \rightarrow p_3(v)|_{m=0} = b_1 v$

- $m=1$    $p_3(v) = \left(\frac{D_1}{5}\right)^{2/5}\frac{1}{\Gamma\left(\frac{2}{5}\right)}\frac{7}{25}(35 + D_1^2 v^5)v \rightarrow p_3(v)|_{m=1} = b_1 v\left(1 + j\frac{Dv^5}{35}\right)$

- $m=2$    $p_3(v) = \left(\frac{D_1}{5}\right)^{2/5}\frac{1}{\Gamma\left(\frac{2}{5}\right)}\frac{7}{5}\frac{12}{1250}(4200 + 120 D_1^2 v^5 + D_1^4 v^{10})v \rightarrow$

  $p_3(v)|_{m=2} = b_1 v\left(1 + j\frac{Dv^5}{35} - \frac{D^2 v^{10}}{4200}\right)$

where $b_1 = \left(\frac{D_1}{5}\right)^{2/5}\frac{1}{\Gamma\left(\frac{2}{5}\right)}$ and $D_1 = \sqrt[4]{-1}\sqrt{D}$. The mathematical property $\Gamma(z+1) = z \cdot \Gamma(z)$ has been also used in the derivation.

$$\boxed{n_3(v) = I_{-\frac{2}{5}}\left(\frac{2}{5}D_1 v^{5/2}\right) = \sum_{m=0}^{\infty} \frac{1}{\Gamma\left(m+\frac{3}{5}\right)m!}\left(\frac{D_1 v^{5/2}}{5}\right)^{2m-\frac{2}{5}}} \qquad (C.2)$$

- $m=0$    $n_3(v) = \left(\frac{D_1}{5}\right)^{-2/5}\frac{1}{\Gamma\left(\frac{3}{5}\right)}\frac{1}{v} \rightarrow n_3(v)|_{m=0} = \frac{b_2}{v}$

- $m=1$    $n_3(v) = \left(\frac{D_1}{5}\right)^{-2/5}\frac{1}{\Gamma\left(\frac{3}{5}\right)}\frac{3}{5}\frac{1}{25}(15 + D_1^2 v^5)\frac{1}{v} \rightarrow n_3(v)|_{m=1} = \frac{b_2}{v}\left(1 + j\frac{Dv^5}{15}\right)$

- $m = 2$
$$n_3(v) = \left(\frac{D_1}{5}\right)^{-2/5} \frac{1}{\Gamma\left(\frac{3}{5}\right)} \frac{3}{5} \frac{8}{5} \frac{1}{1250} (1200 + 80 D_1^2 v^5 + D_1^4 v^{10}) \frac{1}{v} \rightarrow$$

$$n_3(v)|_{m=2} = \frac{b_2}{v}\left(1 + j\frac{Dv^5}{15} - \frac{D^2 v^{10}}{1200}\right)$$

where $b_2 = \left(\frac{D_1}{5}\right)^{-2/5} \frac{1}{\Gamma\left(\frac{3}{5}\right)}$ and $D_1 = \sqrt[4]{-1}\sqrt{D}$. The mathematical property $\Gamma(z+1) = z \cdot \Gamma(z)$ has been also used in the derivation.

Next, we have rewritten the general solution in the following way:

$$\boxed{i(v) = k_1 g_1(v) + k_2 g_2(v)} \tag{C.3}$$

where:

$$k_1 = a_1 b_1$$

$$k_2 = a_2 b_2$$

$$g_1(v) = v^2 \left(1 + j\frac{Dv^5}{35} - \frac{D^2 v^{10}}{4200} + \cdots\right)$$

$$g_2(v) = \left(1 + j\frac{Dv^5}{15} - \frac{D^2 v^{10}}{1200} + \cdots\right)$$

where $k_1$ and $k_2$ are complex constants.

If we now differentiate the general solution (C.1) respect to $v$ and equate it with $\frac{di}{dv} = j\omega \frac{\mu W^2 k^2}{2 I_{DS}} v^3 u = D_2 v^3 u$, it yields:

$$\frac{di}{dv} = k_1 \frac{dg_1}{dv} + k_2 \frac{dg_2}{dv} = D_2 v^3 u \rightarrow \boxed{\frac{u}{B} = k_1 F_1(v) + k_2 F_2(v)} \tag{C.4}$$

where $D_2 = jD I_{DS}$; $B = \frac{1}{D_2}$; $F_1(v) = \frac{dg_1}{dv}/v^3$; and $F_2(v) = \frac{dg_2}{dv}/v^3$

- **Case B:** $\frac{k}{C^7}|v| \gg 1$

Equation (B2.4) can be written in the following way

$$i(v) = v\left[\left(\frac{i_d n_1 v_{cs} - i_s n_2 v_{cd}}{v_{cs} v_{cd}(p_2 n_1 - n_2 p_1)}\right) p_3(v) + \left(\frac{i_s p_2 v_{cd} - i_d p_1 v_{cs}}{v_{cs} v_{cd}(p_2 n_1 - n_2 p_1)}\right) n_3(v)\right] = a_1 f_1(v) + a_2 f_2(v)$$

where: $a_1 = \frac{i_d n_1 v_{cs} - i_s n_2 v_{cd}}{v_{cs} v_{cd}(p_2 n_1 - n_2 p_1)}$, $a_2 = \frac{i_s p_2 v_{cd} - i_d p_1 v_{cs}}{v_{cs} v_{cd}(p_2 n_1 - n_2 p_1)}$; $f_1(v) = v \cdot p_3(v)$; and $f_2(v) = v \cdot n_3(v)$.

and:

$$\boxed{p_3(v) = I_{\frac{1}{3}}\left(\frac{1}{3} D_1 v^3\right) = \sum_{m=0}^{\infty} \frac{1}{\Gamma\left(m + \frac{4}{3}\right) m!}\left(\frac{D_1 v^3}{6}\right)^{2m + \frac{1}{3}}} \tag{C.5}$$

- $m = 0$
$$p_3(v) = \left(\frac{D_1}{6}\right)^{1/3} \frac{1}{\Gamma\left(\frac{4}{3}\right)} v \rightarrow p_3(v)|_{m=0} = b_1 v$$

- $m = 1$
$$p_3(v) = \left(\frac{D_1}{6}\right)^{1/3} \frac{1}{\Gamma\left(\frac{4}{3}\right)} \frac{4}{3} \frac{1}{36}(48 + D_1^2 v^6) v \rightarrow p_3(v)|_{m=1} = b_1 v\left(1 + j\frac{Dv^6}{48}\right)$$

- $m = 2$       $p_3(v) = \left(\frac{D_1}{6}\right)^{1/3} \frac{1}{\Gamma\left(\frac{4}{3}\right)} \frac{4}{3} \frac{7}{2592} (8064 + 168 D_1^2 v^6 + D_1^4 v^{12}) v \rightarrow$

$$p_3(v)|_{m=2} = b_1 v \left(1 + j \frac{D v^6}{48} - \frac{D^2 v^{12}}{8064}\right)$$

where $b_1 = \left(\frac{D_1}{6}\right)^{1/3} \frac{1}{\Gamma\left(\frac{4}{3}\right)}$ and $D_1 = \sqrt[4]{-1}\sqrt{D}$. The mathematical property $\Gamma(z+1) = z \cdot \Gamma(z)$ has been also used in the derivation.

$$\boxed{n_3(v) = I_{-\frac{1}{3}}\left(\frac{1}{3} D_1 v^3\right) = \sum_{m=0}^{\infty} \frac{1}{\Gamma\left(m + \frac{2}{3}\right) m!} \left(\frac{D_1 v^3}{6}\right)^{2m - \frac{1}{3}}} \tag{C.6}$$

- $m = 0$       $n_3(v) = \left(\frac{D_1}{6}\right)^{-1/3} \frac{1}{\Gamma\left(\frac{2}{3}\right)} \frac{1}{v} \rightarrow n_3(v)|_{m=0} = \frac{b_2}{v}$

- $m = 1$       $n_3(v) = \left(\frac{D_1}{6}\right)^{-1/3} \frac{1}{\Gamma\left(\frac{2}{3}\right)} \frac{1}{3} \frac{1}{6} \left(4 + \frac{D_1^2 v^6}{6}\right) \frac{1}{v} \rightarrow n_3(v)|_{m=1} = \frac{b_2}{v} \left(1 + j \frac{D v^6}{24}\right)$

- $m = 2$       $n_3(v) = \left(\frac{D_1}{6}\right)^{-1/3} \frac{1}{\Gamma\left(\frac{2}{3}\right)} \frac{2}{3} \frac{5}{3} \frac{1}{432} \left(480 + 20 D_1^2 v^6 + \frac{D_1^4 v^{12}}{4}\right) \frac{1}{v} \rightarrow$

$$n_3(v)|_{m=2} = \frac{b_2}{v} \left(1 + j \frac{D v^6}{24} - \frac{D^2 v^{12}}{1920}\right)$$

where $b_2 = \left(\frac{D_1}{6}\right)^{-1/3} \frac{1}{\Gamma\left(\frac{2}{3}\right)}$ and $D_1 = \sqrt[4]{-1}\sqrt{D}$. The following property for the Gamma function: $\Gamma(z+1) = z \cdot \Gamma(z)$ is used.

Next, we have rewritten the general solution in the following way:

$$\boxed{i(v) = k_1 g_1(v) + k_2 g_2(v)} \tag{C.7}$$

where:

$$k_1 = a_1 b_1$$
$$k_2 = a_2 b_2$$
$$g_1(v) = v^2 \left(1 + j \frac{D v^6}{48} - \frac{D^2 v^{12}}{8064} + \cdots\right)$$
$$g_2(v) = \left(1 + j \frac{D v^6}{24} - \frac{D^2 v^{12}}{1920} + \cdots\right)$$

where $k_1$ and $k_2$ are complex constants.

If we now differentiate the general solution respect to $v$ and equate it with $\frac{di}{dv} = j\omega \frac{\mu W^2 k^3}{2C' I_{DS}} \text{Sign}[v] v^4 u = D_2 v^4 u$, it yields:

$$\frac{di}{dv} = k_1 \frac{dg_1}{dv} + k_2 \frac{dg_2}{dv} = D_2 v^4 u \rightarrow \boxed{\frac{u}{B} = k_1 F_1(v) + k_2 F_2(v)} \tag{C.8}$$

where $D_2 = jDI_{DS}$; $B = 1/D_2$; $F_1(v) = \frac{dg_1}{dv}/v^4$; and $F_2(v) = \frac{dg_2}{dv}/v^4$.

Now, applicable for both cases, the following short-circuit admittance parameters are calculated for the two-port network depicted in Figure S2:

$$\begin{pmatrix} i_1 \\ i_2 \end{pmatrix} = \begin{pmatrix} y_{11} & y_{12} \\ y_{21} & y_{22} \end{pmatrix} \begin{pmatrix} v_1 \\ v_2 \end{pmatrix}$$

$$y_{11} = \frac{i_1}{v_1}\Big|_{v_2=0} \qquad y_{12} = \frac{i_1}{v_2}\Big|_{v_1=0} \qquad y_{21} = \frac{i_2}{v_1}\Big|_{v_2=0} \qquad y_{22} = \frac{i_2}{v_2}\Big|_{v_1=0}$$

**Figure S2.-** Equivalent circuit in a general form of a FET in two-port configuration described by the short-circuit admittance parameters.

In doing so, the following formulas are used:

$$i(v_{cx}) = k_1 g_1(v_{cx}) + k_2 g_2(v_{cx})$$

$$\frac{u_{cx}}{B} = k_1 F_1(v_{cx}) + k_2 F_2(v_{cx})$$

$$V_C(x) = v_{cx} = \frac{C' - \sqrt{C'^2 \pm 2k[C_t'(V_G - V_{G0} - V_X) + C_b'(V_B - V_{B0} - V_X)]}}{\pm k}$$

$$V_c(x,\omega) = u_{cx} = \frac{C_t'(V_x - V_g) + C_b'(V_x - V_b)}{\sqrt{C'^2 \pm 2k[C_t'(V_G - V_{G0} - V_X) + C_b'(V_B - V_{B0} - V_X)]}} = h_G(V_X)(V_x - V_g) + h_B(V_X)(V_x - V_b)$$

$$u_{cx} = h_{GX}(V_x - V_g) + h_{BX}(V_x - V_b) \tag{C.9}$$

where:

$$\begin{cases} h_G(V_X) = h_{GX} = \dfrac{C_t'}{\sqrt{C'^2 \pm 2k[C_t'(V_G - V_{G0} - V_X) + C_b'(V_B - V_{B0} - V_X)]}} \\ h_B(V_X) = h_{BX} = \dfrac{C_b'}{C_t'} h_{GX} \end{cases} \tag{C.10}$$

and the subscript $X$ stands for drain ($D$) and source ($S$).

So, considering that the source and back-gate are AC short-circuited, thus, $V_s = V_b = 0$:

$$\begin{cases} u_{cs} = -h_{GS}V_g \\ u_{cd} = h_{GD}(V_d - V_g) + h_{BD}V_d \end{cases}$$

- **Calculation of the input admittance $y_{11}$ and the forward transadmittance $y_{21}$**

If the drain is AC short-circuited to the source ($v_2 = V_d - V_s = 0 \rightarrow V_d = V_s$), and a small-signal sinusoidal voltage $v_1 = V_g - V_s$ is applied between gate and source, then:

At the source: $\begin{cases} V_X = V_S \\ V_X = V_s = 0 \end{cases} \rightarrow \begin{cases} v_{cs} \\ u_{cs} = -h_{GS}v_1 \end{cases}$ 　　　　At the drain: $\begin{cases} V_X = V_D \\ V_X = V_d = 0 \end{cases} \rightarrow \begin{cases} v_{cd} \\ u_{cd} = -h_{GD}v_1 \end{cases}$

$$\begin{cases} \dfrac{u_{cs}}{B} = k_1 F_1(v_{cs}) + k_2 F_2(v_{cs}) \\ \dfrac{u_{cd}}{B} = k_1 F_1(v_{cd}) + k_2 F_2(v_{cd}) \end{cases} \rightarrow \begin{cases} k_1 = \dfrac{F_{2s}h_{GD} - F_{2d}h_{GS}}{B(F_{1s}F_{2d} - F_{1d}F_{2s})} v_1 \\ k_2 = \dfrac{F_{1d}h_{GS} - F_{1s}h_{GD}}{B(F_{1s}F_{2d} - F_{1d}F_{2s})} v_1 \end{cases}$$

where: $F_{1s} = F_1(v_{cs})$, $F_{1d} = F_1(v_{cd})$, $F_{2s} = F_2(v_{cs})$, $F_{2d} = F_2(v_{cd})$.

The currents of both ports are calculated as:

$$i_2 = i_d = i(v_{cd}) = k_1 g_1(v_{cd}) + k_2 g_2(v_{cd}) = k_1 g_{1d} + k_2 g_{2d}$$

$$i_s = -i(v_{cs}) = -k_1 g_{1s} - k_2 g_{2s}$$

$$i_1 = i_g = -\frac{C_t'}{C'}(i_d - i_s) = \gamma[k_1(g_{1s} - g_{1d}) + k_2(g_{2s} - g_{2d})]$$

where $\gamma = \dfrac{C_t'}{C'}$, $g_{1s} = g_1(v_{cs})$, $g_{1d} = g_1(v_{cd})$, $g_{2s} = g_2(v_{cs})$, $g_{2d} = g_2(v_{cd})$.

Finally:

$$\boxed{y_{11} = \left.\frac{i_1}{v_1}\right|_{v_2=0} = \gamma \frac{(F_{2s}h_{GD} - F_{2d}h_{GS})(g_{1s} - g_{1d}) + (F_{1d}h_{GS} - F_{1s}h_{GD})(g_{2s} - g_{2d})}{B(F_{1s}F_{2d} - F_{1d}F_{2s})}}$$

$$\boxed{y_{21} = \left.\frac{i_2}{v_1}\right|_{v_2=0} = \frac{(F_{2s}h_{GD} - F_{2d}h_{GS})g_{1d} + (F_{1d}h_{GS} - F_{1s}h_{GD})g_{2d}}{B(F_{1s}F_{2d} - F_{1d}F_{2s})}}$$

(C.11)

- **Calculation of the output admittance $y_{22}$ and the reverse transfer admittance $y_{12}$**

If the gate is AC short-circuited to the source ($v_1 = V_g - V_s = 0 \rightarrow V_g = V_s$), and a small-signal sinusoidal voltage $v_2 = V_d - V_s$ is applied between drain and source, then:

At the source: $\begin{cases} V_X = V_S \\ V_x = V_s = 0 \end{cases} \rightarrow \begin{cases} v_{cs} \\ u_{cs} = 0 \end{cases}$       At the drain: $\begin{cases} V_X = V_D \\ V_x = V_d = v_2 \end{cases} \rightarrow \begin{cases} v_{cd} \\ u_{cd} = (h_{GD} + h_{BD})v_2 \end{cases}$

$$\begin{cases} 0 = k_1 F_1(v_{cs}) + k_2 F_2(v_{cs}) \\ \dfrac{u_{cd}}{B} = k_1 F_1(v_{cd}) + k_2 F_2(v_{cd}) \end{cases} \rightarrow \begin{cases} k_1 = \dfrac{-F_{2s}(h_{GD} + h_{BD})}{B(F_{1s}F_{2d} - F_{1d}F_{2s})} v_2 \\ k_2 = \dfrac{F_{1s}(h_{GD} + h_{BD})}{B(F_{1s}F_{2d} - F_{1d}F_{2s})} v_2 \end{cases}$$

We calculate the currents as:

$$i_2 = i_d = i(v_{cd}) = k_1 g_1(v_{cd}) + k_2 g_2(v_{cd}) = k_1 g_{1d} + k_2 g_{2d}$$

$$i_s = -i(v_{cs}) = -k_1 g_{1s} - k_2 g_{2s}$$

$$i_1 = i_g = -\frac{C_t'}{C'}(i_d - i_s) = \gamma[k_1(g_{1s} - g_{1d}) + k_2(g_{2s} - g_{2d})]$$

Finally:

$$\boxed{y_{12} = \left.\frac{i_1}{v_2}\right|_{v_1=0} = h_{GD}\frac{F_{1s}(g_{2s} - g_{2d}) - F_{2s}(g_{1s} - g_{1d})}{B(F_{1s}F_{2d} - F_{1d}F_{2s})}}$$

$$\boxed{y_{22} = \left.\frac{i_2}{v_2}\right|_{v_1=0} = (h_{GD} + h_{BD})\frac{F_{1s}g_{2d} - F_{2s}g_{1d}}{B(F_{1s}F_{2d} - F_{1d}F_{2s})}}$$

(C.12)

where we have used the following property $\gamma = h_{GD}/(h_{GD} + h_{BD})$.

## C1.   Zero-order approximation ($\omega \to 0$)

To build a zero-order NQS model, the terms in $\omega$ are neglected in the admittance parameter calculation, i.e., we apply the limit $\omega \to 0$. Therefore, the functions $g_1(v)$, $g_2(v)$, $F_1(v)$ and $F_2(v)$ are computed only up to $m = 0$, resulting in:

**Case A:** $\frac{k}{C'}|v| \ll 1$ 

$g_1(v) = v^2$

$g_2(v) = 1$

$F_1(v) = \frac{2}{v^2}$

$F_2(v) = 0$

**Case B:** $\frac{k}{C'}|v| \gg 1$

$g_1(v) = v^2$

$g_2(v) = 1$

$F_1(v) = \frac{2}{v^3}$

$F_2(v) = 0$

Consequently, the admittance parameters calculated according to (C.11) and (C.12) are given by:

- Input admittance $y_{11}$: $\boxed{y_{11} = 0}$
- Reverse transfer admittance $y_{12}$: $\boxed{y_{12} = 0}$
- Forward transadmittance $y_{21}$: $\boxed{y_{21} = g_{m0}}$
- Output admittance $y_{22}$: $\boxed{y_{22} = g_{ds0}}$

**Case A:** $\frac{k}{C'}|v| \ll 1$

$$g_{m0} = -3I_{DS}\frac{h_{GD}v_{cd}{}^2 - h_{GS}v_{cs}{}^2}{v_{cd}{}^3 - v_{cs}{}^3} \rightarrow \boxed{g_{m0} = -\mu\frac{W}{L}\frac{k}{2}(h_{GD}v_{cd}{}^2 - h_{GS}v_{cs}{}^2)}$$

$$g_{ds0} = 3\frac{h_{GD}}{\gamma}I_0\frac{v_{cd}{}^2}{v_{cd}{}^3 - v_{cs}{}^3} \rightarrow \boxed{g_{ds0} = \mu\frac{W}{L}\frac{k}{2}(h_{GD} + h_{BD})v_{cd}{}^2}$$

**Case B:** $\frac{k}{C'}|v| \gg 1$

$$g_{m0} = -4I_{DS}\frac{h_{GD}v_{cd}{}^3 - h_{GS}v_{cs}{}^3}{v_{cd}{}^4 - v_{cs}{}^4} \rightarrow \boxed{g_{m0} = -\mu\frac{W}{L}\frac{k^2}{2C'}(\text{Sign}[v_{cd}]h_{GD}v_{cd}{}^3 - \text{Sign}[v_{cs}]h_{GS}v_{cs}{}^3)}$$

$$g_{ds0} = 4\frac{h_{GD}}{\gamma}I_{DS}\frac{v_{cd}{}^3}{v_{cd}{}^4 - v_{cs}{}^4} \rightarrow \boxed{g_{ds0} = \mu\frac{W}{L}\frac{k^2}{2C'}(h_{GD} + h_{BD})|v_{cd}|^3}$$

We have used the following expressions for the DC current $I_{DS}$, after evaluating (A1.4):

**Case A:** $\frac{k}{C'}|v| \ll 1$

$$I_{DS} = \mu\frac{W}{L}\frac{k}{6}(v_{cd}{}^3 - v_{cs}{}^3)$$

**Case B:** $\frac{k}{C'}|v| \gg 1$

$$I_{DS} = \mu\frac{W}{L}\frac{k^2}{8C'}\left(\text{Sign}[v_{cd}]v_{cd}{}^4 - \text{Sign}[v_{cs}]v_{cs}{}^4\right)$$

We can go from the equivalent circuit depicted in Figure S2 to the one shown in Figure S3 by calculating the new admittances as follows:

$$\begin{cases} y_1 = y_{11} + y_{12} \rightarrow \boxed{y_1 = 0} \\ y_2 = -y_{12} \rightarrow \boxed{y_2 = 0} \\ y_m = y_{21} - y_{12} \rightarrow \boxed{y_m = g_{m0}} \\ y_0 = y_{22} + y_{12} \rightarrow \boxed{y_0 = g_{ds0}} \end{cases} \quad\quad (C1.1)$$

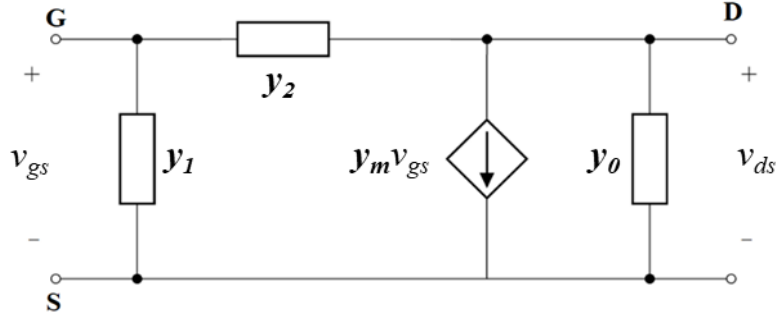

**Figure S3** Equivalent circuit in a standard form of a FET in two-port configuration described by admittance parameters.

## C2. First-order approximation and equivalent circuit

To build a first-order NQS model, second-order and higher terms in $\omega$ are neglected in the admittance parameter calculation. Therefore, the functions $g_1(v)$, $g_2(v)$, $F_1(v)$ and $F_2(v)$ are computed only up to $m = 1$, resulting in:

**Case A:** $\frac{k}{C'}|v| \ll 1$

$$g_1(v) = v^2 \left(1 + j\frac{Dv^5}{35}\right)$$

$$g_2(v) = 1 + j\frac{Dv^5}{15}$$

$$F_1(v) = \frac{2}{v^2} + j\frac{Dv^3}{5}$$

$$F_2(v) = j\frac{Dv}{3}$$

**Case B:** $\frac{k}{C'}|v| \gg 1$

$$g_1(v) = v^2 \left(1 + j\frac{Dv^6}{48}\right)$$

$$g_2(v) = 1 + j\frac{Dv^6}{24}$$

$$F_1(v) = \frac{2}{v^3} + j\frac{Dv^3}{6}$$

$$F_2(v) = j\frac{Dv}{4}$$

Consequently, the admittance parameters calculated according to (C.11) and (C.12) are given by:

Input admittance: $\boxed{y_{11} = j\omega g_{m0}\gamma\frac{\tau_4}{1+j\omega\tau_1}}$      (C2.1)

Reverse transfer admittance: $\boxed{y_{12} = -j\omega g_{m0}\frac{\tau_2}{1+j\omega\tau_1}}$      (C2.2)

Forward transadmittance: $\boxed{y_{21} = g_{m0}\frac{1-j\omega\tau_2}{1+j\omega\tau_1}}$      (C2.3)

Output admittance: $\boxed{y_{22} = g_{ds0}\frac{1+j\omega\tau_3}{1+j\omega\tau_1}}$      (C2.4)

where:

**Case A:** $\frac{k}{C'}|v| \ll 1$

$$g_{m0} = -3I_{DS}\frac{h_{GD}v_{cd}^2 - h_{GS}v_{cs}^2}{v_{cd}^3 - v_{cs}^3} \rightarrow \boxed{g_{m0} = -\mu\frac{W}{L}\frac{k}{2}(h_{GD}v_{cd}^2 - h_{GS}v_{cs}^2)}$$

$$g_{ds0} = 3\frac{h_{GD}}{\gamma}I_0\frac{v_{cd}^2}{v_{cd}^3 - v_{cs}^3} \rightarrow \boxed{g_{ds0} = \mu\frac{W}{L}\frac{k}{2}(h_{GD} + h_{BD})v_{cd}^2}$$

$$\tau_1 = -\frac{D'}{10}\frac{v_{cd}^3 v_{cs}^3(v_{cd} + v_{cs})}{v_{cd}^2 + v_{cd}v_{cs} + v_{cs}^2}$$

$$\tau_2 = \frac{D'}{30} \frac{(2v_{cd}{}^5 - 5v_{cd}{}^2 v_{cs}{}^3 + 3v_{cs}{}^5)}{(h_{GD}v_{cd}{}^2 - h_{GS}v_{cs}{}^2)} h_{GD}v_{cd}{}^2$$

$$\tau_3 = -\frac{D'}{30} (2v_{cd}{}^5 - 5v_{cd}{}^2 v_{cs}{}^3 + 3v_{cs}{}^5)$$

$$\tau_4 = \frac{D'}{30} \frac{[h_{GS}v_{cs}{}^2 (3v_{cd}{}^5 - 5v_{cd}{}^3 v_{cs}{}^2 + 2v_{cs}{}^5) + h_{GD}v_{cd}{}^2 (3v_{cs}{}^5 - 5v_{cs}{}^3 v_{cd}{}^2 + 2v_{cd}{}^5)]}{h_{GD}v_{cd}{}^2 - h_{GS}v_{cs}{}^2}$$

**Case B:** $\frac{k}{C'}|v| \gg 1$

$$g_{m0} = -4I_{DS} \frac{h_{GD}v_{cd}{}^3 - h_{GS}v_{cs}{}^3}{v_{cd}{}^4 - v_{cs}{}^4} \rightarrow \boxed{g_{m0} = -\mu \frac{W}{L} \frac{k^2}{2C'} (\text{Sign}[v_{cd}] h_{GD}v_{cd}{}^3 - \text{Sign}[v_{cs}] h_{GS}v_{cs}{}^3)}$$

$$g_{ds0} = 4 \frac{h_{GD}}{\gamma} I_{DS} \frac{v_{cd}{}^3}{v_{cd}{}^4 - v_{cs}{}^4} \rightarrow \boxed{g_{ds0} = \mu \frac{W}{L} \frac{k^2}{2C'} (h_{GD} + h_{BD})|v_{cd}|^3}$$

$$\tau_1 = -\frac{D'}{12} \frac{v_{cd}{}^4 v_{cs}{}^4}{v_{cd}{}^2 + v_{cs}{}^2}$$

$$\tau_2 = \frac{D'}{24} \frac{(v_{cd}{}^6 - 3v_{cd}{}^2 v_{cs}{}^4 + 2v_{cs}{}^6)}{(h_{GD}v_{cd}{}^3 - h_{GS}v_{cs}{}^3)} h_{GD}v_{cd}{}^3$$

$$\tau_3 = -\frac{D'}{24} (v_{cd}{}^6 - 3v_{cd}{}^2 v_{cs}{}^4 + 2v_{cs}{}^6)$$

$$\tau_4 = \frac{D'}{24} \frac{(v_{cd}{}^2 - v_{cs}{}^2)^2}{(h_{GD}v_{cd}{}^3 - h_{GS}v_{cs}{}^3)} [h_{GD}v_{cd}{}^3 (v_{cd}{}^2 + 2v_{cs}{}^2) + h_{GS}v_{cs}{}^3 (2v_{cd}{}^2 + v_{cs}{}^2)]$$

and $D' = \frac{D}{\omega}$, we have used the following expressions for the DC current $I_{DS}$, after evaluating (A1.4):

**Case A:** $\frac{k}{C'}|v| \ll 1$
$$I_{DS} = \mu \frac{W}{L} \frac{k}{6} (v_{cd}{}^3 - v_{cs}{}^3)$$

**Case B:** $\frac{k}{C'}|v| \gg 1$
$$I_{DS} = \mu \frac{W}{L} \frac{k^2}{8C'} (\text{Sign}[v_{cd}] v_{cd}{}^4 - \text{Sign}[v_{cs}] v_{cs}{}^4)$$

We can go from the equivalent circuit depicted in Figure S2 to the one shown in Figure S3 by calculating the new admittances as follows:

$$\begin{cases} y_1 = y_{11} + y_{12} \rightarrow \boxed{\boldsymbol{y_1 = j\omega g_{m0} \frac{(\gamma \tau_4 - \tau_2)}{1 + j\omega \tau_1}}} \\[2mm] y_2 = -y_{12} \rightarrow \boxed{\boldsymbol{y_2 = j\omega g_{m0} \frac{\tau_2}{1 + j\omega \tau_1}}} \\[2mm] y_m = y_{21} - y_{12} \rightarrow \boxed{\boldsymbol{y_m = \frac{g_{m0}}{1 + j\omega \tau_1}}} \\[2mm] y_0 = y_{22} + y_{12} \rightarrow \boxed{\boldsymbol{y_0 = g_{ds0} \frac{(1 + j\omega \tau_3 [1 - \gamma])}{1 + j\omega \tau_1}}} \end{cases} \quad \text{(C2.5)}$$

where we have used the following relation: $g_{m0}\tau_2 = \tau_3 \gamma g_{ds0}$. (C1.5) yields an equivalent circuit in which the gate-source and gate-drain admittances, $y_1$ and $y_2$ respectively, are simple *RC* networks, and the drain-source is formed by the parallel combination of a frequency dependent current source $y_m v_{gs}$ and the output admittance which is a simple *LC* network. Figure S4 shows the first-order NQS equivalent circuit of a GFET based on lumped elements, which in turn result:

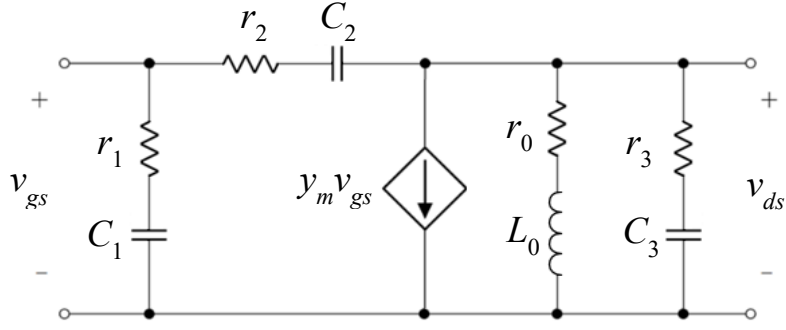

**Figure S4** First-order NQS equivalent circuit of a GFET in two-port configuration described by lumped elements.

where the elements are:

$$C_1 = g_{m0}(\gamma\tau_4 - \tau_2)$$

$$r_1 = \frac{\tau_1}{C_1}$$

$$C_2 = g_{m0}\tau_2$$

$$r_2 = \frac{\tau_1}{C_2}$$

$$y_m = \frac{g_{m0}}{1 + j\omega\tau_1}$$

$$r_0 = \frac{1}{g_{ds0}}$$

$$L_0 = \frac{\tau_1}{g_{ds0}}$$

$$C_3 = g_{ds0}\tau_3(1 - \gamma)$$

$$r_3 = \frac{\tau_1}{C_3}$$

Specifically:

**Case A:** $\frac{k}{C'}|v| \ll 1$

$$C_1 = \frac{3}{10}WLk\frac{(\mathrm{Sign}[v_{cd}]h_{GD}v_{cd}^2)(2v_{cd}^3 + 4v_{cd}^2v_{cs} + 6v_{cd}v_{cs}^2 + 3v_{cs}^3)(-1+\gamma) + (\mathrm{Sign}[v_{cs}]h_{GS}v_{cs}^2)(3v_{cd}^3 + 6v_{cd}^2v_{cs} + 4v_{cd}v_{cs}^2 + 2v_{cs}^3)\gamma}{(v_{cd}^2 + v_{cd}v_{cs} + v_{cs}^2)^2}$$

$$r_1 = \frac{6L}{\mu Wk}\frac{v_{cd}^3v_{cs}^3(v_{cd} + v_{cs})}{(v_{cd} - v_{cs})^2(v_{cd}^2 + v_{cd}v_{cs} + v_{cs}^2)}\frac{1}{(h_{GD}v_{cd}^2)(2v_{cd}^3 + 4v_{cd}^2v_{cs} + 6v_{cd}v_{cs}^2 + 3v_{cs}^3)(-1+\gamma) + (h_{GS}v_{cs}^2)(3v_{cd}^3 + 6v_{cd}^2v_{cs} + 4v_{cd}v_{cs}^2 + 2v_{cs}^3)\gamma}$$

$$C_2 = \frac{3}{10}WLk\frac{\mathrm{Sign}[v_{cd}]h_{GD}v_{cd}^2(2v_{cd}^5 - 5v_{cd}^2v_{cs}^3 + 3v_{cs}^5)}{(v_{cd}^3 - v_{cs}^3)^2}$$

$$r_2 = \frac{6L}{\mu Wk}\frac{v_{cd}^3v_{cs}^3(v_{cd} + v_{cs})}{(v_{cd} - v_{cs})^2(v_{cd}^2 + v_{cd}v_{cs} + v_{cs}^2)}\frac{1}{(h_{GD}v_{cd}^2)(2v_{cd}^3 + 4v_{cd}^2v_{cs} + 6v_{cd}v_{cs}^2 + 3v_{cs}^3)}$$

$$L_0 = \frac{18}{5}\frac{L^3\gamma}{\mu^2 Wk}\frac{v_{cd}^3v_{cs}^3(v_{cd} + v_{cs})}{(v_{cd}^2 + v_{cd}v_{cs} + v_{cs}^2)^3(v_{cd} - v_{cs})^2}\frac{1}{(\mathrm{Sign}[v_{cd}]h_{GD}v_{cd}^2)}$$

$$r_0 = \frac{2L\gamma}{\mu Wkh_{GD}v_{cd}^2}$$

$$C_3 = \frac{3}{10}WLk\frac{\mathrm{Sign}[v_{cd}]h_{GD}v_{cd}^2(2v_{cd}^5 - 5v_{cd}^2v_{cs}^3 + 3v_{cs}^5)}{(v_{cd}^3 - v_{cs}^3)^2}\left(\frac{1}{\gamma} - 1\right) = C_2\left(\frac{1}{\gamma} - 1\right)$$

$$r_3 = \frac{6L}{\mu Wk}\frac{v_{cd}^3v_{cs}^3(v_{cd} + v_{cs})}{(v_{cd} - v_{cs})^2(v_{cd}^2 + v_{cd}v_{cs} + v_{cs}^2)}\frac{1}{(h_{GD}v_{cd}^2)(2v_{cd}^3 + 4v_{cd}^2v_{cs} + 6v_{cd}v_{cs}^2 + 3v_{cs}^3)}\left(\frac{\gamma}{1 - \gamma}\right) = r_2\left(\frac{\gamma}{1 - \gamma}\right)$$

**Case B:** $\frac{k}{C'}|v| \gg 1$

$$C_1 = \frac{2}{3}WLk\frac{(\text{Sign}[v_{cd}]h_{GD}v_{cd}^3)(v_{cd}^2 + 2v_{cs}^2)(-1+\gamma) + (\text{Sign}[v_{cs}]h_{GS}v_{cs}^3)(2v_{cd}^2 + v_{cs}^2)\gamma}{(v_{cd}^2 + v_{cs}^2)^2}$$

$$r_1 = \frac{4C'L}{\mu Wk^2}\frac{v_{cd}^4 v_{cs}^4}{(v_{cd}^2 + v_{cs}^2)(v_{cd}^2 - v_{cs}^2)^2}\frac{1}{(\text{Sign}[v_{cd}]h_{GD}v_{cd}^3)(v_{cd}^2 + 2v_{cs}^2)(-1+\gamma) + (\text{Sign}[v_{cs}]h_{GS}v_{cs}^3)(2v_{cd}^2 + v_{cs}^2)\gamma}$$

$$C_2 = \frac{2}{3}WLk\frac{\text{Sign}[v_{cd}]h_{GD}v_{cd}^3(v_{cd}^2 + 2v_{cs}^2)}{(v_{cd}^2 + v_{cs}^2)^2}$$

$$r_2 = \frac{4C'L}{\mu Wk^2}\frac{v_{cd}^4 v_{cs}^4}{(v_{cd}^2 - v_{cs}^2)^2}\frac{1}{(\text{Sign}[v_{cd}]h_{GD}v_{cd}^3)(v_{cd}^4 + 3v_{cd}^2 v_{cs}^2 + 2v_{cs}^4)}$$

$$L_0 = \frac{16}{3}\frac{C'^2 L^3 \gamma}{\mu^2 Wk^3}\frac{v_{cd}^4 v_{cs}^4}{(v_{cd}^2 + v_{cs}^2)(v_{cd}^2 - v_{cs}^2)^2}\frac{1}{(\text{Sign}[v_{cd}]h_{GD}v_{cd}^3)}$$

$$r_0 = \frac{2C'L\gamma}{\mu Wk^2}\frac{1}{(\text{Sign}[v_{cd}]h_{GD}v_{cd}^3)}$$

$$C_3 = \frac{2}{3}WLk\frac{\text{Sign}[v_{cd}]h_{GD}v_{cd}^3(v_{cd}^2 + 2v_{cs}^2)}{(v_{cd}^2 + v_{cs}^2)^2}\left(\frac{1}{\gamma}-1\right) = C_2\left(\frac{1}{\gamma}-1\right)$$

$$r_3 = \frac{4C'L}{\mu Wk^2}\frac{v_{cd}^4 v_{cs}^4}{(v_{cd}^2 - v_{cs}^2)^2}\frac{1}{(\text{Sign}[v_{cd}]h_{GD}v_{cd}^3)(v_{cd}^4 + 3v_{cd}^2 v_{cs}^2 + 2v_{cs}^4)}\left(\frac{\gamma}{1-\gamma}\right) = r_2\left(\frac{\gamma}{1-\gamma}\right)$$